\documentclass{vldb}
\long\def\comment#1{}

\pdfoutput=1

\usepackage{url}

\usepackage{booktabs} % For formal tables
\usepackage{epsfig}
\usepackage{graphicx}
\usepackage{epstopdf}
\usepackage{amsmath}
\usepackage{xspace}
\usepackage{color}
\usepackage{multirow}
\usepackage{xcolor,colortbl}
\usepackage{weiwAlgorithm}
\usepackage[noend]{algorithmic}
\usepackage{balance}
\usepackage{subfigure}

\usepackage{graphics}

\makeatletter
\newcommand{\nosemic}{\renewcommand{\@endalgocfline}{\relax}}% Drop semi-colon ;
\newcommand{\dosemic}{\renewcommand{\@endalgocfline}{\algocf@endline}}% Reinstate semi-colon ;
\newcommand{\pushline}{\Indp}% Indent
\newcommand{\popline}{\Indm\dosemic}% Undent
\makeatother

\definecolor{Gray}{gray}{0.8}

\newcounter{example}[section]
\renewcommand{\theexample}{\nthesection.\arabic{example}}
\newenvironment{example}{
     \refstepcounter{example}
     {\vspace{1ex} \noindent\bf  Example  \theexample:}}{
     \eop\vspace{1ex}} %\hspace*{\fill}\vspace*{1ex}}

\newcommand{\nthesection}{\arabic{section}}
\newcommand{\eop}{\hspace*{\fill}\mbox{$\Box$}}

\newcommand{\cfig}{Figure~}
\newcommand{\ctab}{Table~}
\newcommand{\csec}{Section~}

\newcommand{\calg}{Algorithm~}

\newcommand{\cexa}{\text{Example~}}

\newcommand{\etc}{etc.}

\newcommand{\stitle}[1]{\vspace{1ex} \noindent{\bf #1}}

\newcommand{\kw}[1]{{\ensuremath {\mathsf{#1}}}\xspace}

\newcommand{\bsp}{\kw{BSP}}
\newcommand{\prg}{\kw{Pregel}}
\newcommand{\prgs}{\kw{PRG}}
\newcommand{\ppl}{\kw{Pregel+}}
\newcommand{\ppls}{\kw{PPL}}

\newcommand{\blg}{\kw{Blogel}}
\newcommand{\blgs}{\kw{BLG}}

\newcommand{\pg}{\kw{PowerGraph}}
\newcommand{\pgs}{\kw{PG}}
\newcommand{\newg}{\kw{Graph3S}}
\newcommand{\newgs}{\kw{G3S}}

\newcommand{\graphd}{\kw{GraphD}}
\newcommand{\gd}{\kw{GD}}
\newcommand{\gdir}{\kw{GDIR}}

\newcommand{\dblp}{\kw{DBLP}}
\newcommand{\orkut}{\kw{Orkut}}
\newcommand{\twit}{\kw{Twitter}}
\newcommand{\uk}{\kw{UK}}
\newcommand{\friend}{\kw{Friendster}}
\newcommand{\clueweb}{\kw{ClueWeb}}
\newcommand{\dbs}{\kw{DB}}
\newcommand{\ors}{\kw{OR}}
\newcommand{\tws}{\kw{TW}}
\newcommand{\frs}{\kw{FR}}
\newcommand{\cws}{\kw{CW}}

\newcommand{\newm}{\kw{TLNE}}
\newcommand{\lani}{\kw{LANI}}
\newcommand{\tav}{\kw{TLV}}
\newcommand{\tas}{\kw{TLS}}

\newcommand{\sunit}{\kw{CU}}

\newcommand{\sep}{seperable\xspace}
\newcommand{\nonsep}{non-seperable\xspace}

\newcommand{\expr}{NE}
\newcommand{\vtype}{V}

\newcommand{\maxlen}{\kw{INT\_MAX}}

\newcommand{\sssp}{\kw{SSSP}}
\newcommand{\bfs}{\kw{BFS}}
\newcommand{\tric}{\kw{TC}}
\newcommand{\tricf}{\kw{Triangle\ Counting}}
\newcommand{\ccomp}{\kw{CC}}
\newcommand{\ccompf}{\kw{Connected\ Component}}
\newcommand{\pr}{\kw{PR}}
\newcommand{\prf}{\kw{Pagerank}}
\newcommand{\prstop}{\kw{10}}
\newcommand{\ppr}{\kw{PPR}}
\newcommand{\core}{\kw{Core}}

\newcommand{\clr}{\kw{Color}}
\newcommand{\clrf}{\kw{Graph\ Coloring}}
\newcommand{\mis}{\kw{MIS}}
\newcommand{\mm}{\kw{MM}}

\vldbTitle{Graph3S: A Simple, Speedy and Scalable Distributed Graph Processing System}
\vldbAuthors{Xubo Wang, Lu Qin, Lijun Chang, Ying Zhang, Dong Wen, and Xuemin Lin}
\vldbDOI{https://doi.org/TBD}
\vldbVolume{13}
\vldbNumber{xxx}
\vldbYear{2020}

\begin{document}

% ****************** TITLE ****************************************
%\title{Simple is Better: A User Friendly Distributed Graph Computing System on Big Graphs}
%\title{Program Like a Vertex: A User Friendly Distributed Graph Computing System on Big Graphs}
%\title{Program Like an Expression: A Subgraph-based Distributed Graph Computing System on Big Graphs}
%\title{Think Like A Neighbourhood Expression: A User-friendly Distributed Graph Computing System}
%\title{S\textbf{\fontsize{17.28}{20}\textsuperscript{3}}-Graph: A Simple, Speedy and Scalable Distributed Graph Processing System}
\title{Graph3S: A Simple, Speedy and Scalable Distributed Graph Processing System}
%\title{From Think-Like-A-Vertex To Think-Like-A-Neighbourhood-Expression}
%\title{Simplify Distributed Graph Processing: From Think Like A Vertex To Think Like A Neighbourhood Expression}

% possible, but not really needed or used for PVLDB:
%\subtitle{[Extended Abstract]
%\titlenote{A full version of this paper is available as\textit{Author's Guide to Preparing ACM SIG Proceedings Using \LaTeX$2_\epsilon$\ and BibTeX} at \texttt{www.acm.org/eaddress.htm}}}

% ****************** AUTHORS **************************************

\numberofauthors{6} 

\author{
\alignauthor
Xubo Wang\\
       \affaddr{The University of Sydney, Australia}\\
       \email{xubo.wang@sydney.edu.au}
\alignauthor
Lu Qin\\
       \affaddr{University of Technology Sydney, Australia}\\
       \email{lu.qin@uts.edu.au}
% 3rd. author
\alignauthor Lijun Chang\\
       \affaddr{The University of Sydney, Australia}\\
       \email{lijun.chang@sydney.edu.au}
\and  % use '\and' if you need 'another row' of author names
% 4th. author
\alignauthor Ying Zhang\\
       \affaddr{University of Technology Sydney, Australia}\\
       \email{ying.zhang@uts.edu.au}
% 5th. author
\alignauthor Dong Wen\\
       \affaddr{University of Technology Sydney, Australia}\\
       \email{dong.wen@uts.edu.au}
% 6th. author
\alignauthor  Xuemin Lin\\
       \affaddr{University of New South Wales, Australia}\\
       \email{lxue@cse.unsw.edu.au}
}

%\author{
%\alignauthor Xubo Wang$^\star$$^\dagger$, Lu Qin$^\dagger$, Lijun Chang$^\star$, Ying Zhang$^\dagger$, Dong Wen$^\dagger$, and Xuemin Lin$^{\ddagger}$
%\vspace{0.1cm} \\
%       \affaddr{$^\star$The University of Sydney, Australia}\\
%       \affaddr{$^\ddagger$University of Technology Sydney, Australia} \vspace{0.1cm} \\
%       \affaddr{$^\dagger$University of New South Wales, Australia}\\
%	   \email{\affaddr{$^\star$\{xubo.wang,lijun.chang\}@sydney.edu.au, 
%	   $^\dagger$\{lu.qin,ying.zhang,dong.wen\}@uts.edu.au,
%	   $^\ddagger${lxue@cse.unsw.edu.au}}}
%}

\maketitle

\begin{abstract}
Graph is a ubiquitous structure in many domains. The rapidly increasing data volume calls for efficient and scalable graph data processing. In recent years, designing distributed graph processing systems has been an increasingly important area to fulfil the demands of processing big graphs in a distributed environment. Though a variety of distributed graph processing systems have been developed, very little attention has been paid to achieving a good combinational system performance in terms of usage simplicity, efficiency and scalability. To contribute to the study of distributed graph processing system, this work tries to fill this gap by designing a simple, speedy and scalable system. Our observation is that enforcing the communication flexibility of a system leads to the gains of both system efficiency and scalability as well as simple usage. We realize our idea in a system \newg and
conduct extensive experiments with diverse algorithms over big graphs from different domains to test its performance. The results show that, besides simple usage, our system has outstanding performance over various graph algorithms and can even reach up to two orders of magnitude speedup over existing in-memory systems when applying to some algorithms. Also, its scalability is competitive to disk-based systems and even better when less machines are used.
\end{abstract} 

%\vspace*{-0.08cm}
\section{Introduction}

% why need distributed graph processing model
%(1) graph analytics is important ( one ref might be Gatner report )
Graph is a ubiquitous structure representing entities and their relationships. It is applied in many areas including social network, web graph, road network, biology and so on. Basic graph problems like Pagerank, connected component, graph coloring, \etc,  are playing fundamental roles in many real-life applications. Efficiently processing graph data is essential in both research and practice.
% (2) graph engine is important
With the dramatic increasing data volume, a lot of research interests have been shown on designing distributed graph processing systems to process big graphs in a distributed environment\cite{dathathri2018gluon,dean2008mapreduce,fan2017parallelizing,Link:giraph,low2012distributed,malewicz2010pregel,yan2014blogel,yan2015effective,zhang2019optimizing}.

\begin{table}[t]
	\centering
	\caption{System Comparison}
	\scriptsize
	\resizebox{\columnwidth}{!} {
		\begin{tabular}{lllll}
			\toprule
			System & Simplicity & Efficiency & Scalability & Flexibility \\

			\toprule			
			\prg & $\star$ $\star$ $\star$  & $\star$ $\star$ $\star$  & $\star$ $\star$ $\star$   & $\star$ $\star$ $\star$ $\star$ $\star$ \\
			
			\ppl & $\star$ $\star$ & $\star$ $\star$ $\star$ & $\star$ $\star$  & $\star$ $\star$ $\star$ $\star$ $\star$ \\
			
			\pg & $\star$ $\star$ $\star$ & $\star$ $\star$ $\star$ & $\star$  & $\star$ $\star$ $\star$ $\star$  \\
			
			\blg & $\star$ & $\star$ $\star$ $\star$ & $\star$ $\star$  & $\star$ $\star$ $\star$ $\star$ $\star$ \\
			
			\graphd & $\star$ $\star$ & $\star$ & $\star$ $\star$ $\star$ $\star$ $\star$ & $\star$ $\star$ $\star$ $\star$ $\star$ \\

			\newg & $\star$ $\star$ $\star$ $\star$ $\star$  & $\star$ $\star$ $\star$ $\star$ $\star$  & $\star$ $\star$ $\star$ $\star$ $\star$  & $\star$ $\star$ $\star$ $\star$\\
			\bottomrule
		\end{tabular}
	}
	\label{tab:system_compare}
\end{table}

%\begin{table}[t]
%	\centering
%	\caption{System Comparison}
%	\scriptsize
%	\resizebox{\columnwidth}{!} {
%		\begin{tabular}{lllll}
%			\toprule
%			System & \scalebox{.9}[1.0]{User-friendliness} & Scalability & Efficiency \\
%			
%			\toprule			
%			\prg & $\star$ $\star$ $\star$ & $\star$ $\star$ $\star$   & $\star$ $\star$ $\star$    \\
%			
%			\ppl & $\star$ $\star$ & $\star$ $\star$ & $\star$ $\star$ $\star$  \\
%			
%			\pg & $\star$ $\star$ $\star$ & $\star$  & $\star$ $\star$ $\star$  \\
%			
%			\blg & $\star$ & $\star$ $\star$ & $\star$ $\star$ $\star$  \\
%			
%			\graphd & $\star$ $\star$ & $\star$ $\star$ $\star$ $\star$ &  $\star$  \\
%			
%			\newg & $\star$ $\star$ $\star$ $\star$ $\star$  & $\star$ $\star$ $\star$ $\star$ $\star$  & $\star$ $\star$ $\star$ $\star$ $\star$  \\
%			\bottomrule
%		\end{tabular}
%		
%	}
%%	{ \raggedright \vspace*{0.1cm} \hspace{0.2cm} $\ast$ User-friendliness \par}
%	\label{tab:system_compare}
%\end{table}

% importance of S S S
% 
Though many distributed graph processing systems have been proposed, we find that not much effort has been put in achieving a good combinational performance in terms of system usage simplicity, efficiency and scalability. 
\ctab \ref{tab:system_compare} 
%shows the scores in terms of each aspect of existing representative systems according to our experimental results. 
shows the scores in terms of each aspect of existing representative systems. 
The scores are relatively given among the systems based on our experimental results. 
Among them, \prg \cite{malewicz2010pregel}, \ppl\cite{yan2015effective}, \pg \cite{gonzalez2012powergraph, low2012distributed} and \graphd \cite{yan2018graphd} are vertex-centric systems. \blg \cite{yan2014blogel} is a block-centric system. All these systems are in-memory systems except \graphd which is an out-of-core system. 
From the table, we can see that none of the existing systems claims a robust performance in the combination of system simplicity, efficiency and scalability, which are all important to end users when processing big graph data.
The unsatisfactory situation is understandable 
since many works focus on improving system efficiency by introducing 
complicated techniques with more APIs to be implemented by users. 
%which weakens usage simplicity. %
For example, \ppl implements Google's popular model \prg with message reduction and load balancing techniques to improve system efficiency. However, the proposed two modes of \ppl require users to 
not only have the related knowledge to choose between them for different applications and graphs so as to achieve best performance, 
but also implement corresponding APIs in different modes. Hence, from system usage perspective, \ppl is more complicated than \prg.
On the other hand, some studies aim at improving system scalability but usually sacrifices system efficiency.
For instance, \graphd aims to improve scalability on top of \ppl by adopting a semi-streaming model. However, the system efficiency becomes weak because many disk accesses are involved. Even though some techniques like ID recoding are proposed to compensate the efficiency sacrifice, not only more APIs are needed and the usage simplicity is thus decreased, but also the technique is only applicable to certain kinds of algorithms. Moreover, its efficiency is still not comparable to in-memory systems.
As a result, to design a distributed graph processing system that is simple to use and with no compromises in efficiency and scalability is challenging and yet to be studied. In this paper, we aim to tackle this problem
by designing a \textbf{S}imple, \textbf{S}peedy and \textbf{S}calable distributed graph processing system, named \newg.

\begin{figure}[t]
\begin{center}
      	\includegraphics[width=\columnwidth]{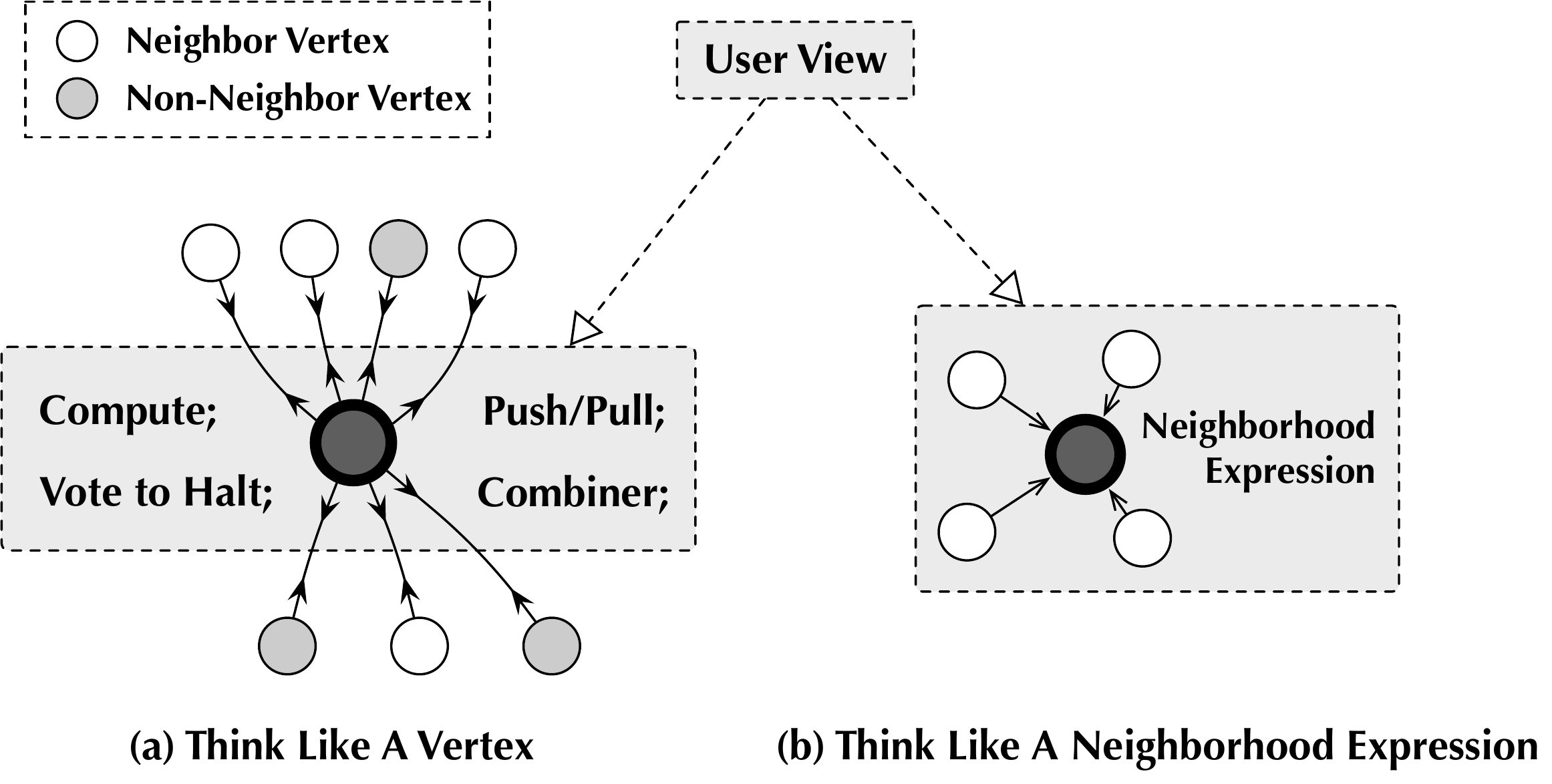}
\end{center}
\caption{Flexibility $\Rightarrow$ \{Simplicity + Efficiency + Scalability\}}
\label{fig:userview}
\end{figure}

%\textcolor{red}{
%	1. flexibility vs. simplicity. why is flexibility important?
%}
%\textcolor{blue}{
%	By flexibility, we mean our system only supports graph processing algorithms where a vertex just communicates with its neighbors. For example, \bfs.
%	However, existing systems has no limitation. They support algorithms where a vertex can communicate with any other vertex. For example, minimum spanning forest algorithm discussed in \cite{yan2015effective}. 
%	In this work, we leave these algorithms out of our scope.
%}
%
%\textcolor{red}{
%	2. current work support neighbor and non-neibors, so it's a superset. why more complicated? just use the subset operations.
%	flexibility
%}
%\textcolor{blue}{
%	Because of widely algorithm supporting nature (better flexibility) of existing systems, they need to expose related APIs to users. 
%	While with our enforcement, \newg manages to only expose computation function to users leading to simple usage.
%}

% limit to neighbor is good
\stitle{Observation.}
%Our main idea is to trade communication flexibility of a system for all three of simplicity, efficiency and scalability. By communication flexibility, we refer to the set of vertices that a vertex in a graph can communicate with. 
We find that a vertex in existing systems is free to communicate with any vertex in a given graph including its neighbor and non-neighbor vertices as shown in \cfig \ref{fig:userview}(a). We call the set of vertices that a vertex can communicate with in a system as the $communication$ $flexibility$ of the system.
Our main idea is to trade the communication flexibility of a system for all three of simplicity, efficiency and scalability.
%As shown in \cfig \ref{fig:userview}(a), a vertex in existing systems is allowed to communicate with any vertex including neighbors and non-neighbors. 
In particular, we propose to enforce the communication range of a vertex in \newg just as its neighborhood as shown in \cfig \ref{fig:userview}(b). 
This enforcement weakens the flexibility of our system, which means \newg only supports graph processing algorithms where a vertex just communicates with its neighbors like Pagerank, connected component, graph coloring and k-core decomposition. 
While existing systems also support algorithms where a vertex communicates with any other vertex like minimum spanning forest algorithm discussed in \cite{yan2015effective}. 
However, our enforcement is acceptable because we can implement all current benchmark algorithms evaluated by the existing distributed graph processing systems. 
%More importantly, based on our enforcement, we propose to save all neighbors of a vertex locally when partitioning a graph into multiple workers. And this idea leads to three-fold benefits. 
More importantly, this enforcement leads to three-fold benefits. 

%\begin{itemize}
%\item
\textit{Easy of programming.} 
The flexible communication in existing systems requires their implementations to expose communication tasks as well as vertex state management to users so that a user can program the communication behaviours of a vertex, 
whereas the enforcement in \newg makes it possible that users only need to provide a computation function based on which a vertex compute its values.
We call this function as a neighborhood expression as shown in \cfig \ref{fig:userview}(b), because it only involves the neighbor vertices.
We name the simple programming model as \textit{Think Like a Neighborhood Expression} (\newm) because users of \newg only provide a neighborhood expression.
%Since a vertex only communicates with its neighbors in our work, it is feasible to 
We save a vertex's neighborhood information locally in a distributed environment. 
Thus, with the neighborhood expression provided by a user, the system can automatically finish all the jobs.
%Therefore, a user only need to design a computation function based on which a vertex computes its value from its neighbor values, and leaves the other jobs to the system. 
%
This will greatly simplify the workload on the user side considering users of existing systems need to take care of not only computation, but also communication, vertex state maintenance, and many optimization realizations (\cfig \ref{fig:userview}(a)). 
\calg \ref{alg:bfs-newg} shows the implementation code of breadth-first search (\bfs) on \newg. The distance $dis$ of a vertex $v$ to source vertex $s$ is updated as the $min\{u.dis+1\}$ where $u$ is a in-neighbor of $v$. 
The neighborhood expression is designated in line~3-4. 
By contrast, in existing systems, a user also needs to design the functions of pushing/pulling messages, when to start/stop computation, combiners and so on \cite{low2012distributed,malewicz2010pregel,yan2014blogel,yan2015effective}.

\begin{algorithm}[t]
	\small
	\caption{BFS on \newg}
	\label{alg:bfs-newg}
	%\algsize
	\KwIn{A given dataset $dataset$; source vertex $s$}
	\KwOut{The distance $V.dis$ of each vertex $V$ to $s$}
	
	\vspace{0.1cm}
	\State {\textbf{Vertex} \vtype(\textbf{int}, dis)}
	\State {\textbf{Graph} G($dataset$)}
	
	\State {G.\textbf{ITER\_N}(V.dis= (ID==$s$) ? 0 : \maxlen, 1)}
	\State {G.\textbf{ITER}(EACH\_IN V.dis=min(V.dis, INB.dis+1))}
\end{algorithm}

%\item
\textit{Efficiency. }
The enforcement also leads to efficient implementations. 
Firstly, different from pushing/pulling required multiple vertex attributes in  existing systems, only changed vertex attributes, we name as $critical$ $attributes$, are synchronized  in \newg which reduces communication cost. 
Secondly, a dual neighbor index is designed to accelerate vertex computation and activation procedure. We also propose a $self$-$adaptive$ $activation$ mechanism based on the dual neighbor index which further improves the efficiency of \newg.
Note that these techniques are not feasible if a vertex communicates with both neighbors and non-neibors considering system scalability. 
%\textit{Speedy}. We propose different techniques to enhance the system efficiency. We introduce the concept of $critical$ $attributes$ where users can designate which attributes to be synchronized for different algorithms so as to reduce communication cost. We also design a dual index structure and a $self$-$adaptive$ $activation$ mechanism which reduce both computation and communication cost for \newg.

%\item
\textit{Scalability. }
Since the applications in \newg only need neighborhood information, we adopt a semi-caching model in \newg where vertex information is saved in memory and neighborhood information is saved on disk. This is practical because vertex values are frequently and randomly accessed and edge information can be scanned linearly.
Besides, edge information usually has a  much higher space cost $O(n^2)$ than that of vertex information $O(n)$ where $n$ is the number of vertices in a given graph.
%\end{itemize}
%\textit{Scalable}. We adopt a semi-caching model for \newg to achieve system scalability. Only vertex values are kept in memory and edge information is stored on disk. This helps improve \newg's scalability because edge information usually has a  much higher space cost than that of vertex information.

Extensive experiments comparing our system with popular state-of-the-art distributed graph processing systems validate 
% the simplicity, speediness and scalability of \newg. 
these benefits of our enforcement in \newg.
Note that the lack of diversity in applications for evaluating system usability and performance is a concern for existing works \cite{kalavri2017high}. Therefore, in this paper, despite commonly used similar and single-stage algorithms, we also include more complex and multi-stage algorithms to test the system performance. We implement $9$ popular graph algorithms on compared systems and evaluate them over six large-scale real datasets from different domains with various characteristics.

\stitle{Contribution. }
In this paper, our principle contributions are shown as follows.

\begin{itemize}
	\item We study the simplicity, efficiency and scalability of a distributed graph processing system by trading communication flexibility.
	\item We design a simple, speedy and scalable system \newg to implement our idea.
	\item We conduct extensive experiments to prove the good performance of our system compared to existing systems.
\end{itemize}

\vspace*{0.03cm}
\stitle{Outline.} The remainder of this paper is organised as follows: \csec \ref{sec:relatedwork} reviews the existing work on graph processing systems. %\csec \ref{sec:model}  introduces our model and 
\csec \ref{sec:realization} introduces our system and implementation techniques.  In \csec \ref{sec:experiments}, extensive experiments over real-life datasets are conducted and results are reported. \csec \ref{sec:conclusion} concludes this paper.

%\vspace*{0.15cm}
\section{Related Work}
\label{sec:relatedwork}

This section reports our review of existing graph processing systems. 
Based on the given graph is processed in a single machine or a cluster, the existing systems could be categorised as either single-machine or distributed systems.

\subsection{Single-Machine (shared-memory) Systems}

Single-machine graph processing systems store and process a given graph in a single machine. 
There are some existing works on single-machine graph processing systems \cite{cheng2015venus,chi2016nxgraph,han2013turbograph,kyrola2012graphchi,nguyen2013lightweight,roy2013x,vora2016load,wang2013asynchronous}.
% in-memory
Ligra \cite{shun2013ligra} adopts a lightweight graph processing framework with two mapping modes: vertex and edge. It dynamically switches between these two modes based on vertex subset density. The framework is efficient on graph traversal algorithms like BFS. 
%
%Galois \cite{kulkarni2007optimistic,nguyen2013lightweight} exploits amorphous data parallelism and supports priority scheduling and dynamic graphs.
% out-memory
GraphChi\cite{kyrola2012graphchi} is a vertex-centric, disk-based system designed for processing large graphs in a single machine. It adopts a method called parallel sliding windows (PSW) for processing large graphs from disks with a very small number of non-sequential accesses to the disk. 
X-Stream\cite{roy2013x} is an edge-centric single-machine system for large-scale graphs by streaming edge data from disk.
%GRACE\cite{} .
TurboGraph \cite{han2013turbograph} is a disk-based graph engine that introduces the pin-and-slide model to perform generalized matrix-vector multiplication on a single machine.
PathGraph \cite{yuan2014fast} is a path-centric graph processing system which partitions a large graph into tree-based partitions and store trees in a DFS order. 
VENUS\cite{cheng2015venus} adopts a vertex-centric streamlined processing model and proposes a new graph storage scheme, v-shards, with two different implementation algorithms.  
FlashGraph \cite{da2015flashgraph} adopts semi-external memory model for graphs stored on fast I/O device like SSD. 
In GridGraph \cite{zhu2015gridgraph}, graphs are partitioned into 1D-partitioned vertex chunks and 2D-partitioned edge blocks. A 2-level hierarchical partitioning is applied to ensure data locality and reduce disk I/O.
NXgraph\cite{chi2016nxgraph} proposes a new structure called Destination-Sorted Sub-Shard to ensure graph data locality and enable fine-grained scheduling. It introduces three updating strategies and adapts to choose the fastest strategy.

Single-machine graph processing systems have high efficiency because of communication cost saving and fast convergence. However, the disadvantage is their weak scalability due to limited hardware sources.
Considering system scalability, in this paper, we aim at a distributed graph processing system which can avoid out-of-memory error by increasing the number of machines until input graph could fit within distributed memory machines.

\subsection{Distributed (shared-nothing) Systems}

For distributed graph processing systems, a given graph is usually partitioned to different machines in a cluster. According to the programming model, existing distributed systems could be divided into vertex-centric (or edge-centric) and subgraph-centric (or block-centric) systems. 
%For each system, users need to be familiar with all the APIs provided to implement an algorithm.

\stitle{Think Like  A Vertex} 
Most existing distributed graph systems adopt the "think like a v
ertex" (\tav) model where users can design an application by specifying the behaviour of a vertex.
Malewick et al. \cite{malewicz2010pregel} first proposed this model and designed a \tav system named \prg which is based on the bulk synchronous parallel (BSP) model \cite{valiant1990bridging}. The BSP model consists of iterations. Inside each iteration, active vertices conduct computation as well as communication with other vertices.
%
%After that, different implementations of \prg are presented. 
Giraph\cite{Link:giraph} is an open-source implementation of \prg in Java. GPS \cite{salihoglu2013gps} presents an optimization technique, large adjacency list partitioning, for high-degree vertices.
Yan et al. \cite{yan2015effective} designed a system named \ppl implementing \prg with message reduction and load balancing techniques.
Zhu et al. \cite{zhu2016gemini} present a distributed system Gemini based on a hybrid push-pull computation model. % 1. dense mode still only for separable operations; 2. mirrors provides only neighbor info but not value, value still needs to be transferred
%
%\stitle{Edge-centric} 
%
GraphLab (PowerGraph) \cite{gonzalez2012powergraph, low2012distributed}  adopts the vertex-cut partition schema and supports both synchronous and asynchronous computation modes. It adopts a Gather, Apply, and Scatter (GAS) programming model where users still think like a vertex. 
%Chaos \cite{roy2015chaos} 

Because distributed in-memory systems provide high efficiency but are weak in scalability, some distributed external-memory systems are proposed to compensate \cite{bu2014pregelix,bu2010haloop,ko2018turbograph++,yan2018graphd}. 
TurboGraph++ \cite{ko2018turbograph++} extends a single out-of-core graph processing system TurboGraph \cite{han2013turbograph} to a distributed environment.  
Yan et al. proposed an out-of-core distributed graph system \graphd \cite{yan2018graphd} based on a semi-streaming model where vertex states are stored in memory and edges and messages are streamed from disk.

There are also some studies on general graph processing system optimization techniques \cite{cai2018efficient, dathathri2018gluon, han2015giraph, liu2017graphene, salihoglu2014optimizing, DBLP:journals/pvldb/SongLWGLJ18, wang2016hybrid, da2015flashgraph, zhong2014medusa}.  
Salihoglu et al. \cite{salihoglu2014optimizing} proposed some optimization techniques to implement algorithms efficiently on \prg-like systems.
Wang et al. \cite{wang2016hybrid} designed an automatic switching mechanism between push and pull computation models to reduce I/O costs on disk data.  
Song et al. \cite{DBLP:journals/pvldb/SongLWGLJ18} put forward a redundancy reduction strategy to achieve high-performance graph analytics by using graph structure. 
The other works focus on improving system efficiency through new hardwares, like SSDs, GPUs\cite{ liu2017graphene, da2015flashgraph, zhong2014medusa}.
%These system independent strategies are considered as a general technique thus not compared in this work. 
We leave these optimization works out of comparison in our study.

\stitle{Think Like A Subgraph} 
There is another category of graph processing systems that allows users to program with a subgraph\cite{chen2018g, fan2017parallelizing, quamar2016nscale, teixeira2015arabesque, tian2013think, yan2014blogel}.
%To overcome the slow information propagation and large communication cost in vertex-like system, some block-based systems are proposed. A given graph is partitioned into different blocks so that vertices inside the same block are more likely connected to each other than vertices in other blocks. Vertices in the same block are partitioned in the same machine and messages can be freely propagated inside a block.  Note that communication inside block won't cause cost. 
%Tian et al. \cite{tian2013think} proposed Giraph++ which opens partition structure to users to program with based on Giraph. 
Yan et al. \cite{yan2014blogel} designed \blg where each connected subgraph is a block and users program functions for blocks.
NScale \cite{quamar2016nscale} and Arabesque \cite{teixeira2015arabesque} adopt the k-hop neighborhood-centric model based on MapReduce framework.  
G-Miner \cite{chen2018g} models subgraph mining problems as independent tasks and provides a task-based pipeline to asynchronously process CPU, Network, Disk I/O operations for efficiency.

%\stitle{Think Like A Subgraph} 
%There is another category of graph processing systems proposed for graph mining problems (e.g., subgraph matching, community detection).  These problems are based on subgraphs, thus subgraph-centric systems is more natural. NScale \cite{quamar2016nscale} and Arabesque \cite{teixeira2015arabesque} adopt k-hop neighborhood-centric model based on MapReduce framework.  G-Miner \cite{chen2018g} models subgraph mining problem as independent tasks and proposes a task-based pipeline to asynchronously process CPU, Network, Disk I/O operations for efficiency.

Despite which model an existing system adopts, users still need to implement multiple functions to specify the behaviours of a vertex or a subgraph. These usually include computation and communication functions, optimization techniques and so on. 
However, as validated by our experiments, our system is not only simple to use but also shows good efficiency and scalability over different algorithms.

%In this section, we first define some basic notations and then discuss related distributed graph computing system.

%\subsection{Pregel}
%\label{sec:pregel}
%
%\prg  is a popular implementation of bulk synchronous parallel (\bsp) \cite{valiant1990bridging}model.
%
%\subsection{Blogel}
%\label{sec:blogel}
%
%\subsection{GraphLab}
%\label{sec:graphlab}
%
%\glb
%
%\subsection{Other Systems}
%\label{sec:othersystem}

%\section{{System Design}}
\section{{The System}}
\label{sec:realization}

To meet the requirement of a good system in terms of simplicity, efficiency and scalability, we design a new system named \newg. In this section, we first provide the overview of \newg, followed by the detailed introduction of proposed techniques about the Simple, Speedy and Scalable aspects of \newg.

\begin{figure}[t]
	\centering
	\includegraphics[width=\columnwidth]{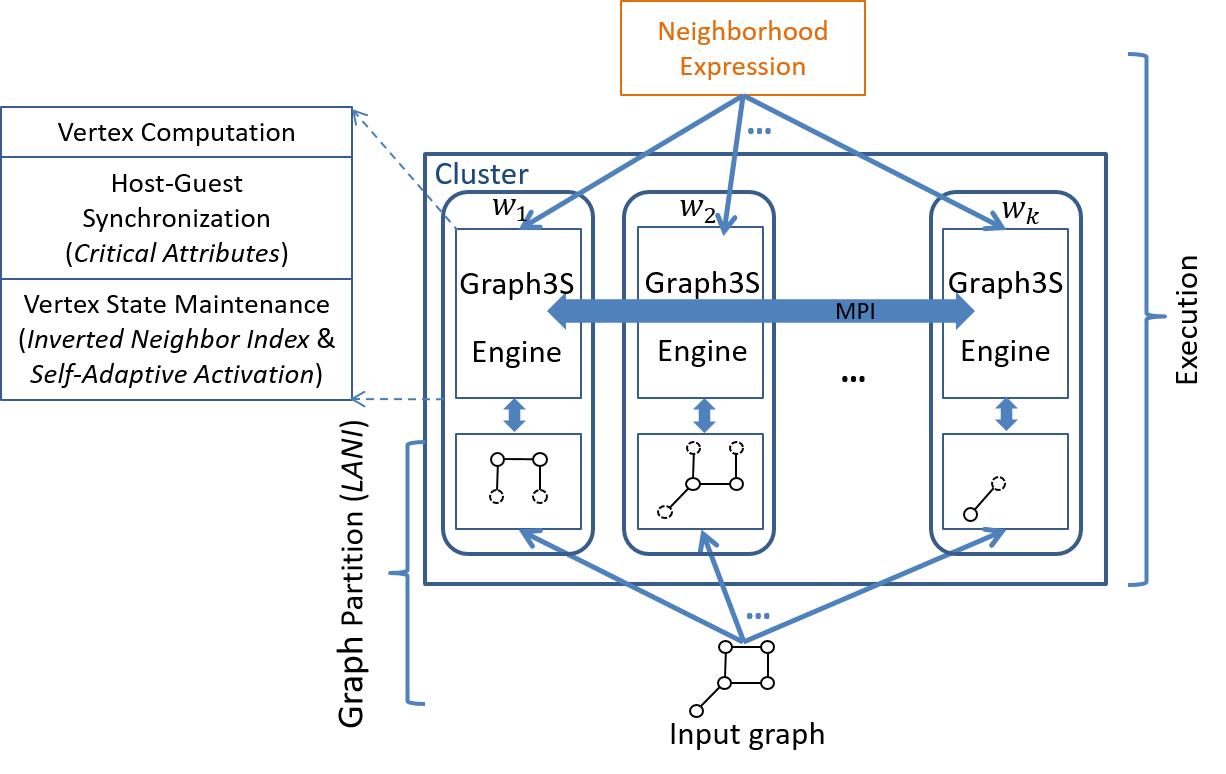}
	%\vspace{-0.1cm}
	\caption{\newg overview.}
	\label{fig:structure}
\end{figure}

\subsection{Overview}
\cfig \ref{fig:structure} shows the overview of \newg. 
A user only needs to provide a neighborhood expression based on which a vertex computes its value to the system. Then the rest of the work is automatically done by \newg.
% structure: 
Input graph is partitioned to different workers in a hashing way. 
We adopt the Bulk Synchronous Parallel (\bsp) model in \newg engine. BSP is based on iterative supersteps (iterations). 
Vertex computation, communication (Host-Guest Synchronization), and vertex state maintenance happen in each superstep with synchronization barrier occurring at the end of each superstep.

\subsection{Simple}
Now we introduce how easily a user can implement an algorithm on \newg. %As we introduced, a user only needs to design the expression based on which a vertex computes its value from its neighborhood information. 

\stitle{Notations} 
%To start the introduction of our model, we firstly introduce some basic notations. 
We use $G=\{V, E\}$ to represent a graph where $V$ and $E$ denote vertex and edge sets respectively. 
%Each vertex $v$ is related with one or more attributes denoted by $attr(v)$. We also use vertex value to vertex attributes interchangebly in the remainder of this paper. 
$G$ is partitioned to a cluster of worker machines $W=\{w_1, w_2, ..., w_k\}$. 
%Here, 
We use $n, m$ and $k$ to represent the numbers of vertices, edges and machines respectively. 
% where $V=\{v_1,...,v_n\}$ and  $E=\{e_1,...,e_m\}$. 
For each $v \in V$, $N(v)$ represents the set of neighbor vertices of $v$. 
%That is $(u,v) \in E$ if $u\in N(v)$.
If $G$ is a directed graph, we use $N_{in}(v)$ and $N_{out}(v)$ to represent in-neighbor and out-neighbor vertex sets respectively.

\subsubsection{{Think Like A Neighborhood Expression}}
\label{sec:model}

%\subsubsection{Simple as An Expression}
\stitle{Simple as An Expression}
We propose to only leave the computation function for users and assign other tasks automatically managed by system. We design a programming model named "think like a neighborhood expression" (\newm) to implement the idea. 
As shown in \cfig \ref{fig:userview}.b, a vertex has a local view of its neighbors. 
A vertex does not need to communicate with other vertices to obtain values it needs.
The neighborhood information is maintained automatically by the system.
Hence, 
the only job for a \newm user is to provide a neighborhood expression $NE$ which tells a vertex how to compute its value. All other jobs are hidden from users and managed automatically by the model. 
This makes developing an application on \newm much easier compared with existing models. 

Note that a vertex in \newm is designed to only communicate with its neighbor vertices and the expression $NE$ is only involved with a vertex's neighbors. 
This design is reasonable because most graph problems can be solved 
with vertex communication within neighborhood range. For example, a vertex $v$ in 
breadth-first search (\bfs) only needs its neighbor vertex distance 
values to compute its own distance value $v.dis$. 
In this case, the neighborhood expression is $v.dis=min_{u\in N_{in}(v)}\{u.dis\}+1$. 
In \prf, a vertex $v$ updates its own ranking value $v.rank$ based on the ranking values of its neighbors. 
The neighborhood expression is $v.rank=(1-a)/n + a*SUM_{u\in N_{in}(v)}(u.rank/|N_{out}(u)|)$ where $a$ is a residual probability constant. 
Similarly, \newm can be applied to many other problems like \ccompf and \tricf.

%\subsubsection{Behind The Scene}
%\label{model-back}

\begin{figure}
	\centering
	\includegraphics[width=0.7\columnwidth]{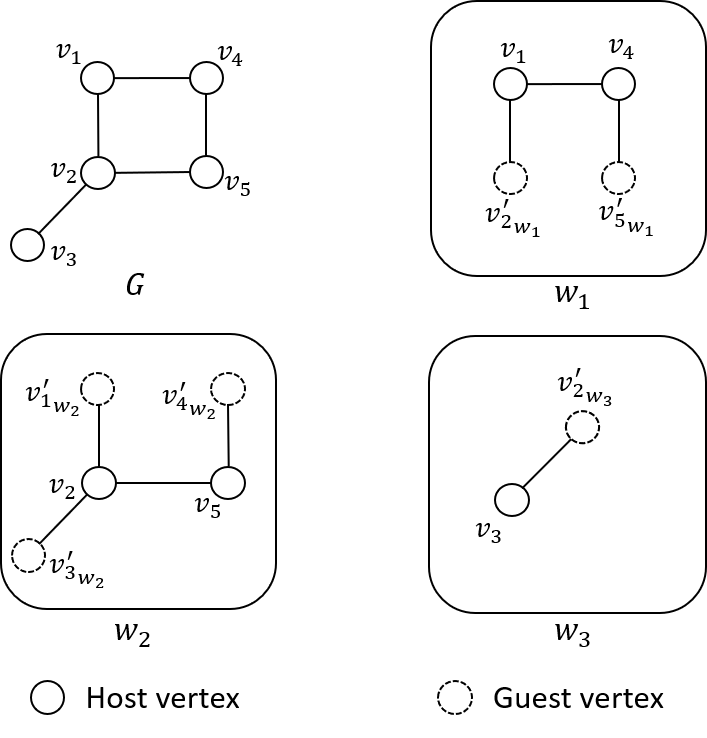}
	%\vspace{-0.1cm}
	\caption{TLNE: locally available neighborhood information example}
	\label{fig:partition}
\end{figure}

\stitle{Locally Available Neighborhood Information} 
The key to the simple usage of \newm is to 
maintain the neighborhood local view of a vertex.
We adopt a mechanism named as Locally Available Neighborhood Information (\lani) to meet the target.

%\lani works like this. 
After partitioning a given graph, 
for those vertices whose neighbors are partitioned to different workers, \lani builds copy vertices of their neighbors  locally.
More specifically, when partitioning a given graph into a distributed environment, suppose that a vertex $v$ is assigned to a worker $w_i$. If any of its neighbor vertices $u\in N(v)$ is partitioned to a different worker $w_j$, a copy vertex $u'_{w_i}$ of $u$ is constructed on $w_i$. 
For ease of expression, we call a vertex that is directly partitioned to a worker as a \textbf{\textit{host}} vertex. The constructed neighbor copy vertices are called as \textbf{\textit{guest}} vertices.
Here, $v$ on $w_i$ and $u$ are both host vertices on $w_j$. 
$u'_{w_i}$ is a guest vertex on $w_i$. 
Note that we call $u'_{w_i}$ on $w_i$ as a corresponding guest vertex for host vertex $u$ on $w_j$.
A host vertex may have more than one corresponding guest vertices because it may be a neighbor of different host vertices on different workers. 
Each host vertex is in charge of synchronizing its corresponding guest vertex values to keep consistency.
In this way, for each host vertex $v$, all its neighborhood information is locally available all the time.

\begin{example}
	\label{example:lani}
	Consider a graph $G$ in \cfig \ref{fig:partition}. All vertices are partitioned to three workers $w_1$, $w_2$ and $w_3$ in the cluster. More specifically, $v_1$ and $v_4$ are on $w_1$, $v_2$ and $v_5$ are on $w_2$ and $v_3$ is on $w_3$. Take vertex $v_1$ as an example. $v_1$ is a host vertex on machine $w_1$. One of its neighbors, vertex $v_4$, is partitioned to the same machine $w_1$ but the other neighbor, $v_2$ is partitioned to a different machine $w_2$. This requires that a guest vertex ${v_2}'_{w_1}$ is constructed in $w_1$ as a copy of vertex $v_2$. So $v_1$ has its neighbor values of $v_2$ and $v_4$ locally at all time. Similarly, for each vertex, all its neighborhood information is locally available and its value can be directly computed based on the given neighborhood expression from users.
\end{example}

It will be much desirable for users if each vertex can see all other vertices locally. However, the space cost will be massive and the model will have very poor scalability. 
It is more reasonable to save the neighborhood information locally. For each worker, the average saving cost is $O(n*d_{avg}/k)$ where $d_{avg}$ is the average vertex degree in a given graph.

\begin{algorithm}[t]
	\small
	\caption{TLNE Execution}
	\label{alg:iteration}
	%\algsize
	
	\vspace{0.1cm}
	\State{ Initialization}
	\For{ (i=1; $Active()\neq \emptyset$ ; i++) }{
		\ForEach{$v\in Active()$} {
			\State{${val(v)}^{i} = NE({N(v)}^{i-1})$}
			\If{$val(v)^{i} \ne val(v)^{i-1}$}{
				\State{ synchronize to all ${v'}$ }
				\State{ activate $v$ and $I(v)$ }
			} 
		} 
	}
\end{algorithm}

%\subsection{Execution}
\stitle{Execution}
Overall, \newm is a synchronous, distributed graph processing model where algorithms run in iterations. 
The pseudo code of \newm's execution procedure is shown in \calg \ref{alg:iteration}.
At the beginning of \newm, a given graph is partitioned to a cluster and guest vertices are constructed. All vertices initialize their values (line~1).
%
%Then computations are conducted in iterations (line~2-7). 
In each iteration, each active host vertex $v$ updates its value $val(v)$ based on the user-provided neighborhood expression $NE$ (line~4). 
${val(v)}^{i}$ represents the value of vertex $v$ in iteration $i$.
Also, if a host vertex's value changes, it synchronizes its new value to corresponding guest vertices represented by $v'$ (line~6). $I(v)$ denotes the set of vertices of which $v$ is a neighbor. In other words, if $val(u)$ is dependent on $val(v)$, then $u\in I(v)$.
%A vertex value needs to be updated if any of its neighbors changes. Thus vertices whose value or neighbor value changes will be activated in the next iteration. 
The model stops computation until no active vertex exists.

\begin{example}
	\label{example:model}
	Take the graph given in \cexa \ref{example:lani} as an example. We use $v_1$ to explain how our model works. Suppose $v_1$ is active in iteration $i$. Then $v_1$ will compute its value based on the user-given neighborhood expression $NE(v_4, {v'_2}_{w_2})$. If $v_1$'s value changes, the new value will be synchronized to all $v_1$'s corresponding guest vertices which is ${v'_1}_{w_2}$ on $w_2$ in this case. Then $v_1$ and its neighbors $v_4$ and $v_2$ are activated in iteration $i+1$. If $val(v_1)$ doesn't change, it remains inactive until it is activated again.
\end{example}

%%Challenges
%\subsubsection{Challenges}
%%\stitle{Challenges} 
%To implement \newm as an efficient and scalable system, several challenges are required to tackle. 
%The first challenge is how to efficiently maintain neighborhood information. 
%The second challenge is how to efficiently implement the automatic vertex activation process managed by system.
%The third challenge is how to provide a good system scalability when saving neighborhood information locally. 
%To address these challenges, we introduce our \textbf{S}imple, \textbf{S}peedy and \textbf{S}calable system named as \newg 
%which not only implements \newm but also has different techniques to guarantee the system efficiency and scalability. 

\subsubsection{Programming API}
\label{sec:api}
The programming APIs of \newg are shown in \ctab \ref{tab:notation}. 
%
%Each application code consists of definitions of vertex, graph and computation. 
The first two APIs are provided for users to easily define vertex attributes and instantiate a graph.
Vertex type is defined by using ${VERTEX\ \vtype(AT1, AN1,AT2,AN2,...)}$ where $AT$ denotes an attribute type and $AN$ represents an attribute name. A vertex value may contain one or more attributes.
A graph instance can be constructed using $GRAPH\ G(input)$. 
Computation is defined by $ITER(\expr)$ or $ITER\_N(\expr)$ for vertex value update. 
$\expr(V, ACCESS\_NB)$ is a neighborhood expression defined by a user where $V$ computes its value by accessing its neighbor values with $ACCESS\_NB$.
Three different ways of $ACCESS\_NB$, namely $EACH\_NB$, $EACH\_IN$ and $EACH\_OUT$ are provided to access neighbors which are neighbor, in-neighbor  and out-neighbor respectively.
If $ITER$ is used, the system will start execution on each active vertex until no active vertex exists. A vertex is activated if its value or any of its neighbor values changes. All these works are automatically maintained by \newg. %We will introduce the automatic mechanism later.
When using $ITER\_N(\expr, n\_itr)$, the system stops running after $n\_itr$ iterations computation.
$attr$ in $ITER(\expr, attr)$ represents the vertex attributes that the system needs to synchronize. We provide this API to improve system efficiency. Details will be discussed in \csec \ref{speedy}.

%\begin{table}[t]
%	\centering
%	\caption{\newg programming API}
%	\scriptsize
%	\resizebox{\columnwidth}{!} {
%		\begin{tabular}{ll}
%			%\hline	
%			\toprule	
%			User Define & $NE(V, ACCESS\_NB)$ \\
%			%\hline
%			
%			\toprule
%			Neighbor Access & ${EACH\_NB, NB}$ \\
%			(${ACCESS\_NB}$) & ${EACH\_IN, INB}$ \\
%			& ${EACH\_OUT, ONB}$ \\
%			%\hline
%			
%			\toprule			
%			System Provide & ${VA(AT1, AN1,AT2,AN2,...)}$ \\ 				
%%			& ${init(NE)}$ \\
%			& ${iter(NE)}$ or $iter\_n(NE,n\_itr)$ \\
%			& ${iter(NE,attr)}$ or $iter\_n(NE,n\_itr, attr)$ \\
%			%\hline
%			\bottomrule
%		\end{tabular}
%	}
%	\label{tab:notation}
%\end{table}

\begin{table}[t]
	\centering
	\caption{\newg programming API}
	\scriptsize
	\resizebox{\columnwidth}{!} {
		\begin{tabular}{ll}
			\toprule
			Definition & API \\

			\toprule
			Vertex & ${VERTEX\ \vtype(AT1, AN1,AT2,AN2,...)}$ \\
			
			%\toprule
			Graph & $GRAPH$ $G{(input)}$ \\
			
			%\toprule			
			Computation  & ${ITER(NE)}$ or $ITER\_N(NE,n\_itr)$  \\
			 & ${ITER(NE,attr)}$ or $ITER\_N(NE,n\_itr, attr)$ \\
			\bottomrule
		\end{tabular}
	}
	\label{tab:notation}
\end{table}

\stitle{BFS}
\calg \ref{alg:bfs-newg} is an example of implementing \bfs on \newg. 
Line~1 defines a vertex type where each vertex has one attribute $dis$ describing the distance between the current and source vertices. 
Line~2 creates a graph from the given graph. 
Line~3 initializes the $dis$ value for each vertex. The $dis$ value of source vertex $s$ is set as 0, and others are set as $INT\_MAX$.
Line~4 gives the neighborhood expression to update the $dis$ value. 
Each vertex $V$ updates its $dis$ value as the minimum value among current value $V.dis$ and $INB.dis+1$ where $INB$ represents V's in-neighbor vertex. $EACH\_IN$ is used here to access all in-neighbor vertices. 
The communication and active vertex maintenance are automatically managed by \newg. The system keeps running until no active vertex remains. 

\begin{algorithm}[!tp]
	\caption{BFS on \ppl}
	\label{alg:bfs-ppl}
	
	\vspace{0.1cm}
	\State{using BFSKeyT=int, BFSMsgT=int}
	
	\Struct{\bf BFSValueT}{
		\State{int dis}
		\State{vector$<$BFSKeyT$>$ out\_nbs}
	}
	\Comment{(De)serialization function omitted}
	\Class{{\bf BFSVertex} : Vertex$<$BFSKeyT, BFSValueT, BFSMsgT$>$ }{
		\Func{void bcast\_to\_out\_nbs()} {
			\State {BFSMsgT msg = value().dis}
			%\State {vector$<$BFSKeyT$>$ nbs=value().out_nbs}
			\ForEach{$onb \in value().out\_nbs$} {
				\State {send\_msg(onb, msg)}
			}
		}
		
		\Func{void compute(MessageContainer $\&$ msgs)} {
			\If{step\_num = 1} {
				\If{id=srcID} {
					\State{value().dis=0}
					\State{ bcast\_to\_out\_nbs()} 
				} \Else{value().dis=\maxlen}
			}\Else{
				\State {value().dis=min(value().dis, min(msgs)+1)}
				\State {bcast\_to\_out\_nbs()}
			}
			\State { vote\_to\_halt()}
		}
	}
	
	\Class{\bf BFSWorker : Worker$<$BFSVertex$>$} {
		\Func{BFSVertex* toVertex(char* line)} {
			\State {char* pch = strtok(line, "\textbackslash t")}
			\State {BFSKeyT id=atoi(pch)}
			\State {BFSVertex* v = new BFSVertex}
			\State {v-$>$id=id; v-$>$value().dis=-1}
			\State { {\bf if} (id==srcID) {\bf then} {v-$>$value().dis=0}; {\bf else} {v-$>$vote\_to\_halt()}}
			\State {int pch\_i=1}
			\While {pch=strtok(NULL, " ")}{
				\If{pch\_i $>$ 1}{
					\State{v-$>$value().out\_nbs.push\_back(atoi(pch))}
				}
				\State{pch\_i++}
			}
			\State {return v}
		}
		%\Comment{output function omitted}
	}
	
	\Class{\bf BFSCombiner : Combiner$<$BFSMsgT$>$}{
		\Func{void combine(BFSMsgT $\&$ old, const BFSMsgT $\&$ new)}{
			\State{ {\bf if} old $>$ new {\bf then} old=new}
		}
	}
	
	\Func{void {\bf bfs\_pregel}(BFSKeyT srcID, string input)}{
		%\State {init_workers()}
		\State {BFSWorker worker}
		\State {BFSCombiner combiner}
		\State {worker.setCombiner($\&$combiner)}
		\State {worker.run(input)}
		%\State {worker\_finalize()}
		%\State {return 0}
	}
\end{algorithm}

The implementation here is much simpler than in existing systems. Because of space limitation, we only give the implementation of \bfs on a popular system \ppl in \calg \ref{alg:bfs-ppl} to show the difference intuitively.
In \ppl, a user needs to define classes of Vertex, Worker, Combiner and so on. For each class, the computation and communication behaviours need to be implemented carefully by users. 
For example, to think like a vertex, besides computations (in line~11-13 and line~15-18), a user also needs to send its value to its neighbors ($bcast\_to\_out\_nbs()$ in line~14 and 19). 
Also, vertex state maintenance need to be managed by user (in line~20).
In addition, a combiner needs to be implemented by users to get better system efficiency (line~34-36). 
These implementations require users to be familiar with many system APIs and decide when to use optimization techniques. 
Note that some details of this implementation are omitted because of space restrictions. Nevertheless, it is obvious that implementations on existing systems are more complicated than that on \newg. 
\stitle{Compare with existing systems} 
As we introduced above, many distributed graph processing systems have been proposed to tackle big graph processing problems. Existing studies focus on improving system efficiency and scalability. 
To the best of our knowledge, the usage simplicity of distributed graph processing systems has not been well discussed yet. 

In the literature, the popular existing graph processing models are "think like a vertex" (\tav)  and "think like a subgraph" (\tas). We call a vertex in \tav or a subgraph in \tas as a computing unit (\sunit). Users of these models design an application by specifying the behaviours of a \sunit. 
%When thinking like a \sunit in these models, a user needs to tell the \sunit 
For instance, how to compute a \sunit's value,  when to start/stop computation, how to get values a \sunit needs and how to send its value to other vertices. 
These involves users in implementing many functions related to computation, communication and \sunit state maintenance. 
For example, in \prg model \cite{malewicz2010pregel}, users need to implement functions in charge of message sending $SendMessageTo(dest\_vertex, message)$, computation and received message processing $Compute(msgs)$ and vertex state maintenance $VoteToHalt()$. %are all required from users. 
Compared to \prg, \ppl requires users to take care of extra APIs related to different modes.
In $GAS$ model of \pg \cite{gonzalez2012powergraph}, at least functions $Gatter$, $Apply$ and $Scatter$ need to be implemented. Function $Gather$ tells a vertex how to get neighbor vertex values. Function $Apply$ combines the gathered values and applies to update its own value. $Scatter$ uses its new value to activate neighbors for next iteration. 
In block-centric system \blg \cite{yan2014blogel}, not only APIs for vertices need to be designed, but also APIs for block computation, communication and state management are also required to be implemented.
We also find that different optimization APIs, like combiners in \prg are provided in existing systems for users to decide when to use them. 
In \graphd, besides basic APIs, extra APIs for ID recoding need to be decided whether applicable and necessary to be implemented.
To efficiently implement an algorithm on existing systems, users need to acquire a clear understanding of the algorithms and professionalism of the system.

%Therefore, our goal is to simplify a user's tasks when developing a distributed graph computing algorithm and also guarantee good efficiency and scalability.

\subsection{Speedy}
\label{speedy}

Adding to its simplicity, we also present the techniques to make our system speedy.

\subsubsection{Host-Guest Synchronization}
%%
%We propose the concept of $critical$ $attributes$ in \newg to speedup host-guest synchronization process.
%
%\stitle{Motivation} 
%% motivation
%We observe that in many applications on existing systems, more \sunit attributes lead to a greater communication cost. This is because a \sunit needs to fetch partial or all attributes from its neighbors to compute its own value. 
%%
%Similarly, in \newg, if a vertex has more than one attribute, synching all attributes is inefficient. 
%\newg should allow users to selectively choose attributes of a vertex to be transferred. In response, we propose the concept of $critical$ $attributes$.

It is easy to understand that more vertex attributes lead to a greater communication cost. If a vertex has more than one attribute, synching all attributes is inefficient. In response, we propose the concept of $critical$ $attributes$ in \newg for users to selectively choose attributes of a vertex to be transferred. 
In this way, the host-guest synchronization process is accelerated.

Note that since the neighborhood information is locally available in \newg, it is reasonable to transfer just partial attributes which need to be updated. However, in many existing systems, a vertex has no local neighborhood information. As a result, they need to send all needed attributes in every computation iteration.

\stitle{Critical Attributes} 
In \newg, two APIs, $ITER(NE, attr)$ and $ITER\_N(NE,n\_itr, attr)$ are provided for users to designate which attributes are to be transferred
when designing an algorithm with multiple attributes vertex.
% $VERTEX$ $\vtype(AT_1, AN_1$$,$$AT_2,AN_2,$$...,AT_p,AN_p)$, 
%a user can use $attr$ in APIs. 
We call these designated attributes as $critical$ $attributes$. 
During host and guest vertex synchronization, \newg only synchronizes the designated critical attributes between host and corresponding guest vertices. In this way, the transformation cost of non-critical attributes are saved.

\begin{algorithm}[t]
	\caption{Coloring on \newg}
	\label{alg:color-newg}
	%\algsize
	\KwIn{A given dataset $dataset$}
	\KwOut{The color $V.color$ of each vertex $V$}
	
	\vspace{0.1cm}	
	\State {\textbf{Vertex} V(\textbf{int}, deg, \textbf{int}, color)}
	\State {\textbf{Graph} G($dataset$)}
	
	\State {\textbf{int} MaxC=INT\_MD}
	\State {\textbf{bool} * used=new bool[MaxC]}
	\State {G.\textbf{ITER\_N}(V.deg=DEG; V.color=-1, 1)}
	
	\State {\nosemic G.\textbf{ITER}	( \,  } 
	\State {\pushline \dosemic memset(used, 0, min(V.deg+1, MaxC)) }
	\State {\nosemic EACH\_NB 	\textbf{if}(NB.deg $>$ V.deg $||$ (NB.deg == V.deg \&\& NB\_ID $>$ ID))\{ }
	\State {\pushline	\textbf{if}(NB.color==-1) \textbf{return}; }
	\State {			used[NB.color]=true; }
	\State {\popline		  \}}
	
	\State {\nosemic \textbf{for}(\textbf{int} i=0; i$<$MaxC; ++i)\{ }
	\State {\pushline	\textbf{if}(!used[i]) \{V.color=i;\textbf{break};\} }
	\State {\popline \dosemic	\} 	, TWO ) \, }
%	\State {G.\textbf{iter}($Min\_Color$, TWO)}

\end{algorithm}

%	\vspace{-0.2cm}	

\begin{example}
	\label{example:color}
%\stitle{\clr} 
Take \clrf (\clr) as an example. \clr is a problem of coloring vertices in a given graph such that no two adjacent vertices share the same color \cite{Link:graphcoloring}. It is a basic graph problem with many practical applications.
In the greedy \clr algorithm, each vertex's color is assigned with the smallest available color that has not been used by its neighbors. 
The vertex order is defined according to vertex degree and ID. A vertex $u$ is larger than $v$ when $u$'s degree is larger than $v$'s. 
We break the tie using vertex ID.
%If they have the same degree, $u$ is larger than $v$ if $u$'s ID is larger. 

\calg \ref{alg:color-newg} shows the implementation of \clr on \newg. 
Each vertex has two attributes: the vertex degree $deg$ and its color value $color$ (line~1). By default, all vertices have an attribute ID in \newg. 
Line~7-14 designate the neighborhood expression for a vertex to update its attribute $color$ based on its neighbor attributes.
The parameter $TWO$ in function $itr$ (line~14) tells the system that the second attribute $color$ is critical. As a result, the system only synchronizes attribute $color$ from host vertices to corresponding guest vertices. 
In existing systems like \prg, a vertex needs to transfer all attributes including itw own ID, degree and color in each iteration so that its neighbor vertices could compute their color values.
In comparison, \newg saves the cost of source ID and degree transformation, giving it greater efficiency.

\end{example}

%\subsubsection{Auto Trigger}
%%\stitle{Auto Trigger} 
%In order to provide system efficiency, we also design an automatic trigger mechanism to efficiently complete vertex state maintenance for \newg. 

\subsubsection{Vertex State Maintenance}
We propose the following methods for efficient vertex state maintenance.

%\subsubsection{Inverted Neighbor Index}
\stitle{Dual Neighbor Index} 
We design a structure called $dual$ $neighbor$ $index$ for each vertex to complete vertex activation efficiently. 
In \newg, a vertex is activated when any of its neighbor values changes. In other words, if a vertex value changes, it activates itself and all vertices of which it is a neighbor. 
To efficiently implement this, we design two indices, neighbor index and inverse neighbor index,
represented by $N(v, w)$ and $I(v, w)$ respectively for each vertex $v$. 
Here, $N(v, w)$ contains vertices that are needed for $v$ to compute its values in worker $w$, namely $v$'s neighbors. 
Note that $N(v, w)$ is only constructed for host vertices in each worker because only host vertices compute their own values from its neighbor values.
$I(v, w)$ includes host vertices in worker $w$ that $v$ needs to notify 
when it updates its values, namely vertices of which $v$ is a neighbor of in worker $w$. Inverse neighbor indices are constructed for each vertex on worker including host and guest vertices.
%.

\begin{example}
	\label{example:index}
Take the graph in \cfig \ref{fig:partition} as an example. 
In worker $w_2$, neighbor indices are built for vertex $v_2$ and $v_5$ and inverse neighbor indices are built for $v_2$, $v_5$, ${v'_1}_{w_2}$, ${v'_3}_{w_2}$ and ${v'_4}_{w_2}$. 
The neighbor index of $v_2$ on worker $w_2$ is
$N(v_2, w_2)=\{{v'_1}_{w_2}, {v'_3}_{w_2}, v_5\}$.
And the inverse neighbor index of $v_2$ is $I(v_2, w_2)=\{v_5\}$.
Besides, the inverse neighbor indices of guest vertex $v'_2$ on other workers $w_1$ and $w_3$ are $I({v'_2}_{w_1}, w_1)=\{v_1\}$ and $I({v'_2}_{w_3}, w_3)=\{v_3\}$ respectively.
\end{example}
%Instead of searching the vertices that a vertex is a neighbor of on-the-fly, we design an inverted neighbor index for each vertex $v$ to save all vertices of which $v$ is a neighbor. %That is $I(v)=\{u:(u,v)\in E$ or $(v,u)\in E\}$. We also use $I(v)$ to represent the inverted neighbor index for vertex $v$.
%If given graph is directed, $I(v)={u:(u,v)\in E or (v,u)\in E}$
%For example, the inverted neighbor index of $v_2$ in \cfig \ref{fig:partition} is $I(v_2)=\{v_1, v_3, v_5\}$.

With the dual index structure, $v$ can use $N(v,w)$ to compute its value efficiently. And if $v$'s value changes, the system can directly activate the vertices in its corresponding inverse neighbor indices. %If no change, then no need to inform. 
Note that these indices are constructed after partition and saved on disk. This is feasible due to our enforcement in this paper. 
A vertex only communicates with its neighbors in our setting, hence the relations could be built offline and can be accessed by linear scanning during execution. 
If a vertex's communication flexibility is as any vertex in a given graph, this would be impractical because the whole graph needs to be saved on each worker.

\begin{figure}[t]
	\includegraphics[width=\columnwidth]{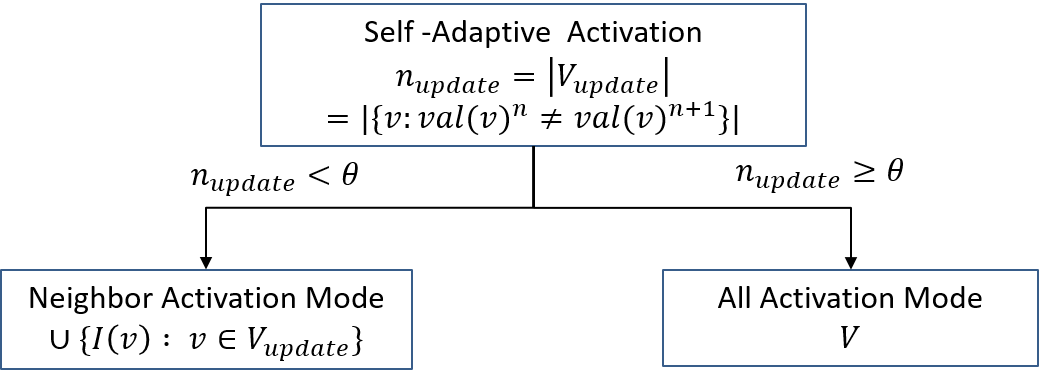}
	\caption{Self-Adaptive Activation.}
	\label{fig:SAA}
\end{figure}

%\subsubsection{Self Adaptive Activation}
\stitle{Self-Adaptive Activation} 
We also propose a self-adaptive activation mechanism to further improve vertex state maintenance efficiency based on the following observation. 

We find that the above activation process works fine when not many vertices change values. However, when most vertices in a given graph update their values, more cost will be spent on obtaining and scanning inverted neighbor indices. 
For example, if $v_2$ and $v_4$ in \cfig \ref{fig:partition} update their values in iteration $n$, then all indices of $v_2$, $v'_2$, $v_4$, and $v'_4$ need to be obtained to know which vertices to activate in iteration $n+1$. In fact, all vertices will be activated in iteration $n+1$ because the union of these index sets equals to $V$. However, if we directly activate all vertices in $G$, the time to scan the dual neighbor indices can be saved.

%Based on our observation,
Thus, instead of obtaining dual neighbor indices and scanning them in every iteration, 
we design two activation modes named $neighbor$ $activation$ $mode$ and $all$ $activation$ $mode$ for \newg 
and the system automatically chooses one of them for activation to get a better performance.
The process of self-adaptive activation is illustrated in \cfig \ref{fig:SAA}. 
$V_{update}=\{v: val(v)^n \ne val(v)^{n+1}\}$ represents the vertices whose values change in current iteration. The total number of updated vertices $n\_change = |V_{update}|$ is recorded during computation in each iteration for the system to adaptively choose an activation mode. 
If $n\_change$ is smaller than a given threshold $\theta$ which means that few vertices change their values, then there is a low probability that the need-to-be-activated vertex set would approximate $V$. In this case, neighbor activation mode is chosen. Each updated vertex obtains its inverted neighbor index and activates the indexed vertices in $\cup\{I(v):v\in V_{update}\}$. 
Otherwise, all activation mode is chosen and all vertices in $V$ are directly activated in the next iteration. In this case, the scanning time of inverted neighbor indices are saved.
Consider running \bfs on \newg. At earlier stage of execution, few vertices change value so neighbor activation mode is chosen. In the middle age, all activation mode is used because most vertices update their values. At the later age, fewer and fewer vertex values change. So the mode goes back to neighbor activation again.

\subsection{Scalable}

As we introduced above, \newm requires a vertex to save all its neighborhood information locally. A direct way to implement \newg would be saving both vertex and edge information in memory. However, this is impractical for big graphs. 
To counter for this, we adopt a semi-caching strategy to enhance the scalability of \newg. 
More specifically, we keep the vertex values in memory but edge information on disk. This is suitable for \newg because of the following three reasons:

\begin{itemize}
	\item Firstly, a vertex only communicates with its neighbors in \newg. During vertex computation, the edge information on disk will only be scanned linearly to get all neighbor information. On the other hand, vertex values are usually accessed randomly and constantly. It is better to save more random and constant accessed information in memory.
	
	\item Secondly, edge information rarely changes in our setting but vertex values are updated frequently. Note that we leave distributed system on dynamic graphs for future work.
	
	\item Thirdly, edge information usually costs more space than vertex information. Given a graph with $n$ vertices, the edge saving cost could be $O(n^2)$ while the vertex information only costs $O(n)$. It is practical to save vertex values in memory.
\end{itemize}

From the above, we can see that the semi-caching model is a good balancing strategy between system efficiency and scalability for \newg. 
Note that the semi-streaming model adopted in \graphd saves both edge information and messages on disk which actually weakens system performance in two aspects. Firstly, message streaming on disk slows down the system efficiency because of more disk accesses involved. Secondly, more disk space are required especially for message intensive algorithms and thus can easily cause out of disk error for real large graphs. These two aspects are both validated in our experiments.

%\vspace*{-0.2cm}
\section{Performance Studies}
\label{sec:experiments}

Our experimental results are outlined herein. 

%info from biggraph
\begin{table}[!t]
	\centering
	\caption{Characteristics of datasets}
	\label{table:dataset}
\resizebox{\columnwidth}{!} {
		\begin{tabular}{lrrrrr}
	    \toprule
		Dataset 	& $|V|$ & $|E|$		& $deg_{max}$	& $deg_{avg}$ 	 \\
        \midrule
		\dbs   & 986,207 &  13,414,472 & 979 & 13.60 \\
		\ors	& 2,997,167 &  212,698,418 &   27,466 & 70.97 \\
		\uk	    & 18,520,343 &  523,574,516 &  194,955	& 28.27 \\ 	
		\tws	& 41,652,230 & 2,936,729,768 &  2,997,487 & 70.51 \\
		\frs & 124,836,180 &  3,612,134,270 &  5,214 & 28.93 \\ 	
		\cws & 978,409,098 & 42,574,107,469	&  75,611,696 & 43.51 \\
        \bottomrule	
	\end{tabular}
}
\end{table}

%info from link
%\begin{table}
%	\centering
%	\caption{Characteristics of datasets}
%	%\scriptsize
%	\begin{tabular}{|l|l|r|r|r|r|r|} \hline
%		\rowcolor{Gray} Dataset 	& Type 	& $|V|$ & $|E|$		& $deg_{max}$	& $deg_{avg}$ 	 \\ \hline
%		\dblp   & u & 986,324 & 6,707,236 &	979 & 6.800  \\ \hline
%		\orkut	& u & 3,072,626 & 234,370,166 &	33,313 & 76.277  \\ \hline	
%		\uk	    & d	& 18,529,486 & 298,113,762 & 194,942	& 16.096 \\ \hline		
%		\twit	& d & 41,652,230 & 	1,468,365,182 & 2,997,469 & 35.253 \\ \hline	
%		\friend & u & 65,608,366 & 	1,806,067,135 &  & 27.528  \\ \hline	
%		\clueweb& d & 978,409,098 & 42,574,107,469	& 75,611,690 & 43.514 \\ \hline	
%	\end{tabular}
%	\label{table:dataset}
%\end{table}

%\stitle{Datasets.} \cite{Poker06} We used 6 real-world datasets of different sizes obtained from LAW \cite{Dataset}. \dblp (\dbs) \cite{Dataset:dblp}, \orkut (\ors) \cite{Dataset:orkut}, \twit (\tws) \cite{Dataset:twitter} and \friend (\frs) \cite{} are social network graphs. \uk \cite{Dataset:uk} and \clueweb (\cws) \cite{Dataset:clueweb} are webgraphs. \ctab \ref{table:dataset} shows the dataset details. 

\stitle{Datasets.} We used 6 real-world datasets of different sizes obtained from LAW \cite{Dataset}. \dblp (\dbs), \orkut (\ors), \twit (\tws) and \friend (\frs) are social network graphs. \uk and \clueweb (\cws) are webgraphs. \ctab \ref{table:dataset} shows the dataset details. 
$|V|$ and $|E|$ represent the number of vertices and edges respectively. $deg_{max}$	and $deg_{avg}$ denote the maximum and average vertex degree in each dataset respectively. %Note that all datasets are processed as undirected graphs in our experiments.

%Road network?

\stitle{Experimental settings.}
We ran our experiments on a cluster of 10 machines, each with one 3.0GHz Intel Xeon E3-1120 CPU (4 cores), 64GB DDR3 RAM and 610GB disk. Unless specified, we use 6 machines, each with 4 cores by default.
%We ran our experiments on a cluster of 4 machines, each with one 2.9GHz Intel Xeon E5-2690 CPU(8 cores) and 32GB DDR3 RAM. Unless specified, the default experiment setting for all systems is 4 machines, each with 2 cores.

We compared our system \newg with the representative systems: vertex-centric \prg, \ppl \cite{System:ppl} and \pg \cite{System:powergraph}, block-centric \blg \cite{System:blogel} and out-of-core system \graphd \cite{System:graphd}. All systems are implemented in C++. 
We use $\theta =n/50$ as the threshold for self-adaptive activation in \newg because it guarantees a good system performance in most cases of our experiments.
We use Yan's implentation \cite{System:ppl} of \prg. 
In terms of \ppl, we adopt the mirroring mode in the experiments. Similar to \cite{System:ppl}, we selected the vertex mirror threshold as the minimum value between 1000 and the value computed using their cost model. 
If not stated, we use the default settings of compared systems.
For ease of expression, we represent the systems \newg, \prg, \ppl, \pg, \blg and \graphd by \newgs, \prgs, \ppls, \pgs, \blgs and \gd respectively in the results.
We also include results of \graphd with ID recoding technique represented by \gdir in the experiments.

% pplg threshold = #machines * e_power(avg_deg/#machines) = 4*e^(a_d/4)

%v[lqin@orion7 run]$ lspci -vvv | grep Ethernet
%00:19.0 Ethernet controller: Intel Corporation 82579LM Gigabit Network Connection (rev 05)
%01:00.0 Ethernet controller: Intel Corporation 82574L Gigabit Network Connection

\stitle{Algorithms.} To evaluate the system performances, we use 9 algorithms 
including single-phase algorithms: Breadth First Search (\bfs), Connected Component (\ccomp), Pagerank (\pr), Personalized Pagerank (\ppr), K-Core Decomposition (\core) \cite{montresor2012distributed} and Graph Coloring (\clr) \cite{linial1992locality} and multi-phase algorithms: Maximal Independent Set (\mis) \cite{luby1986simple}, Maximal Matching (\mm) \cite{linial1992locality} and Triangle Counting (\tric).
%from 3 categories including Traversal style: Breadth First Search (\bfs), Always-Active style: Connected Component (\ccomp), Pagerank (\pr), Personalized Pagerank (\ppr), K-Core Decomposition (\core) \cite{montresor2012distributed} and Graph Coloring (\clr) \cite{linial1992locality}, Multi-Phase style: Maximal Independent Set (\mis) \cite{luby1986simple}, Maximal Matching (\mm) \cite{linial1992locality} and Triangle Counting (\tric). 
%Among them, \bfs is graph traversal algorithms. \pr and \ppr represent always-active graph algorithms. \core and and \clr represent "drop" algorithm which means the number of active nodes drops as time goes. \mis and \mm are multi-phase algorithms.
%
Among them, \bfs, \ccomp, \pr, \ppr and \mis are separable algorithms. \core, \clr and \mm are non-separable algorithms. An algorithm is separable if commutative and associative operation is to be applied on transmitted messages where optimization techniques like combiner can be applied. The ID recoding of \graphd is also only applicable to separable algorithms. 
Details of algorithm implementations are introduced as follows:

\stitle{\bfs.} We implemented \bfs based on \sssp codes from authors of \ppls, \pgs, \blgs and \graphd.

\stitle{\ccomp.} We directly use implementations of \ccomp from authors of compared systems.

\stitle{\pr, \ppr.}  We modified authors' code of \pr from \ppls, \pgs and \graphd by changing the termination condition to \prstop iterations execution. 
%We used uathors' implementation of \pr on \pg setting the termination condition to \prstop iterations computaion. 
For \blg, we used the authors' implementation of \pr and modified the termination condition of second step which operates in V-mode to \prstop iterations computation. We implemented \ppr based on \pr for all systems.

\stitle{\clr.}  
We implemented \clr for both \prg and \ppl. An aggregator is designed to get the number of uncolored vertices and the program terminates when all vertices are colored. No combiner is used since a vertex needs to know the color value of each neighbor and the messages can't be combined. 
\clr is not implemented on \blg because it is \nonsep where the advantage of block model could not be applied. For \sep computing like \bfs and \ccomp, the value of a vertex could be updated continually inside a block. While the color value of a vertex is dependent on all other neighbors' values which means it can't be updated until the next iteration. It follows that \blg needs the same number of iterations as a vertex-centric program and extra time on constructing and maintaining block information. 
We implemented \clr on \pgs and \gd. ID recoding of \gdir is not applicable because \clr is \nonsep.

\stitle{\core.}  Similar to \clr, we implemented \core for both \prgs and \ppls with no combiner. An aggregator is implemented to get the number of vertices whose core values are updated in a current iteration. The algorithm terminates when no core value changes. 
We also implemented \core on \pg. 
\core is not implemented on \blg as computing core value of a vertex is \nonsep. 
\core is not implemented on \gd and \gdir since aggregator is not provided on \graphd and termination condition cannot be implemented.

\stitle{\tric.} We implemented \tric for for both \prgs and \ppls. We adopted authors' implementations of Triangle Counting for \pgs and \gd. ID recoding of \gdir is not applicable because \tric is \nonsep. 
\tric is not implemented on \blg because it is \nonsep.

\stitle{\mis.} We programmed \mis on \prgs, \ppls and \pgs. For \prgs and \pgs, a combiner is designed to combine messages. \mis is not implemented on \blg because the block technique is not effective on multi-phase algorithms. \mis is also implemented on both \gd and \gdir.

\stitle{\mm.} We implemented \mm for for both \prgs, \ppls and \pg. Note that combiner is not applicable for \mm. We also implemented \mm on \pg. Similar to \mis, \mm is not implemented on \blg. \graphd doesnot support \mm. %since the system implementation does not support algorithms where a vertex has more than one attribute besides id and degree.

\stitle{Metrics.} We report the \emph{running time} and \emph{communication cost} to compare the system performances. 
\emph{Running time} is counted from the moment when the data graph is totally loaded in the cluster to the time when the computation is completed. Note that data loading and result dumping time are excluded. 
\emph{Communication cost} is the sum of data size transferred among workers in the cluster.
Note that neither the cost of partitioning an input graph nor distributing it to workers is included.

%\begin{table}[t]
%	\centering
%	\caption{Number of Code Lines}
%	\label{table:usage}
%\resizebox{\columnwidth}{!} {
%		\begin{tabular}{lrrrrrrrr}
%		\toprule
%		System 	& \bfs & \pr	& \ccomp	& \clr & \core & \tric & \mis & \mm	 \\
%		\midrule
%		\newgs   &  15 &  14 	&  14 &  26 &  17 &  23 &  25 &  27  \\
%		\prgs	& 110 & 106 	&  94 & 130 & 121 & 130 & 111 & 115  \\
%		\ppls	&  84 & 109 	&  75 & 125 & 128 & 107 &  97 & 143  \\		
%		\pgs	& 101 &  98	&  99 & 126 &  99 & 311 & 113 & 141  \\		
%		\blgs	& 212 & 319+169 & 159 &  NA &  NA &  NA &  NA &  NA  \\		
%		\gd		&  76 &  61	 	&  74 & 119 &  NA &  104+53 &  93 &  NA   \\	
%		\gdir	&  57 &  45 	&  57 &  NA &  NA &  NA &  74 &  NA  	\\						
%		\bottomrule	
%	\end{tabular}
%}
%\end{table}

\begin{table}[t]
	\centering
	\caption{Number of Code Lines}
	\label{table:usage}
	\resizebox{\columnwidth}{!} {
		\begin{tabular}{lrrrrrrrr}
			\toprule
			System 	& \bfs & \pr	& \ccomp	& \clr & \core & \tric & \mis & \mm	 \\
			\midrule
			\newgs   &  15 &  15 	&  15 &  27 &  18 &  24 &  26 &  28  \\
			\prgs	& 101 & 107 	&  95 & 131 & 122 & 120 & 112 & 116  \\
			\ppls	&  85 &  110	&  76 & 126 & 129 & 108 &  98 & 144  \\		
			\pgs	& 106 &  103	&   & 129 &  102 & 319 & 116 & 144  \\		
			\blgs	& 221 & 328+178 & 16 &  NA &  NA &  NA &  NA &  NA  \\		
			\gd		&  84 &  66	 	&  78 & 123 &  NA &  108+57 &  97 &  NA   \\	
			\gdir	&  62 &  50 	&  62 &  NA &  NA &  NA &  78 &  NA  	\\						
			\bottomrule	
		\end{tabular}
	}
\end{table}

\subsection{Usage Simplicity Comparison}
\label{sec:exp1}

Besides the discussion about the APIs comparison in \csec \ref{sec:api}, we also show the number of code lines  in \ctab \ref{table:usage} using tokei \cite{Tool:tokei} as a reference for simplicity comparison. 
%We format the source codes of different algorithms on compared systems using the code format tool in Eclipse IDE C/C++ and list the number of code lines in \ctab \ref{table:usage}. 
%Note that this may not be an exact answer because different people format codes in different ways but it can still give an intuitive understanding of the implementation complexities of different systems. 
%
From \ctab \ref{table:usage}, we can see that implementations on \newg are much simpler compared with existing systems. 
This is easy to understand because the main task for a \newg user is to give a neighborhood expression for vertex computation. While the existing system users need to implement different APIs in terms of not only computation, but also tasks including communication, vertex state management, and so on.
%For example, a user of \prgs and \ppls needs to implement functions managing vertex computation, message sending/receiving and vertex state maintenance. Combiners and aggregators also need to be considered and implemented to provide better system performance. For \pgs, a user needs to employ Gather, Apply and Scatter functions. For \blgs, computation and communication functions of not only a vertex but also a block need to be implemented. The implementation of \graphd is similar to \ppls.
%
The difference is more severe when developing and tuning multi-phase algorithms like \tric, \mm and \mis because the extra tasks in each phase accumulates for users to take.

\begin{figure*}[t]
	\subfigure[\bfs]{ \label{fig:general_bfs}
		\includegraphics[scale=0.2]{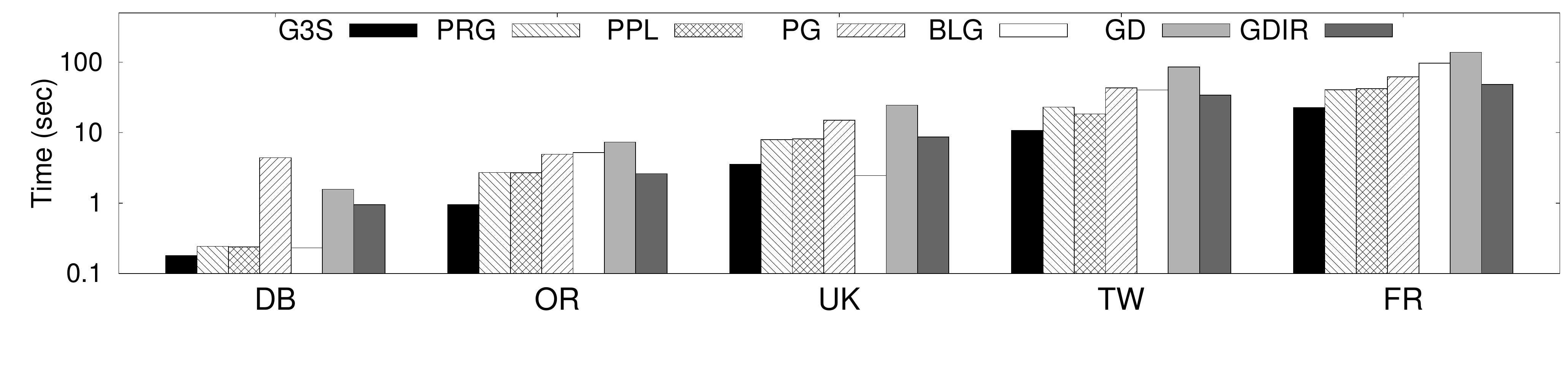}
	}
	\hfill
	\subfigure[\ccomp]{ \label{fig:general_cc}
		\includegraphics[scale=0.2]{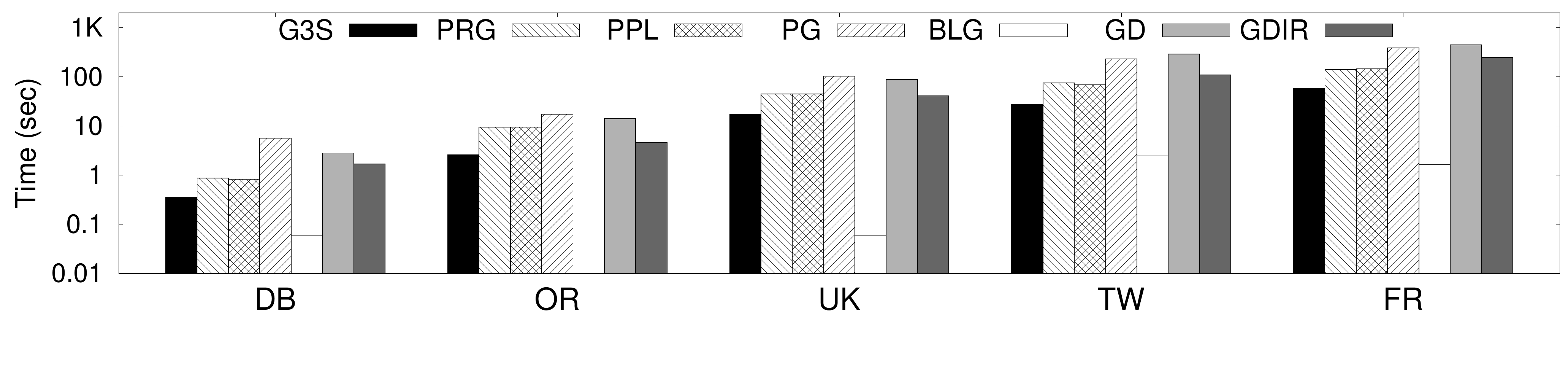}
	}
	\hfill
	\subfigure[\pr]{ \label{fig:general_pr}
		\includegraphics[width=0.485\linewidth, height=2.1cm]{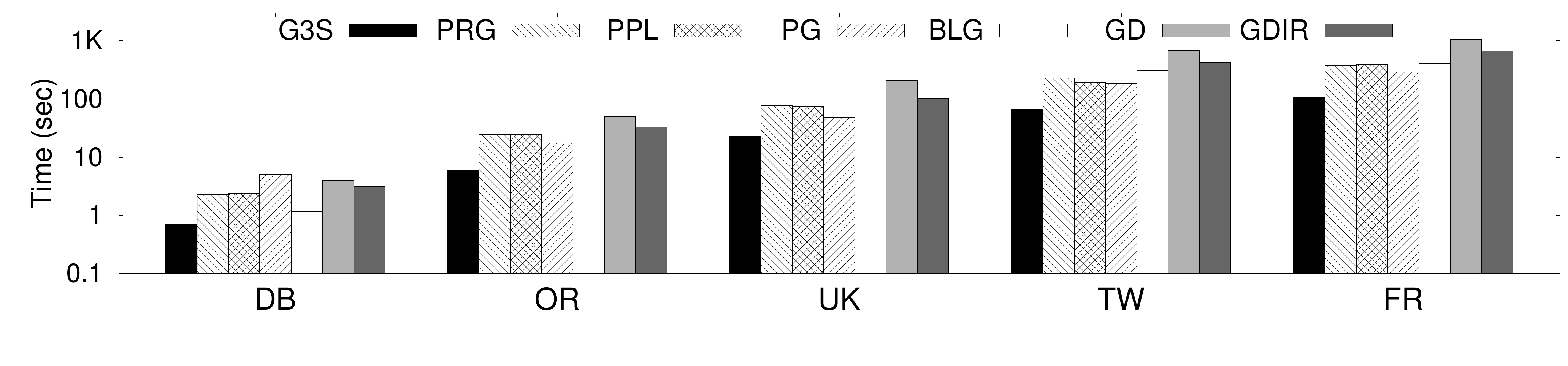}
	}
	\hfill
	\subfigure[\ppr]{ \label{fig:general_ppr}
		\includegraphics[width=0.485\linewidth, height=2.1cm]{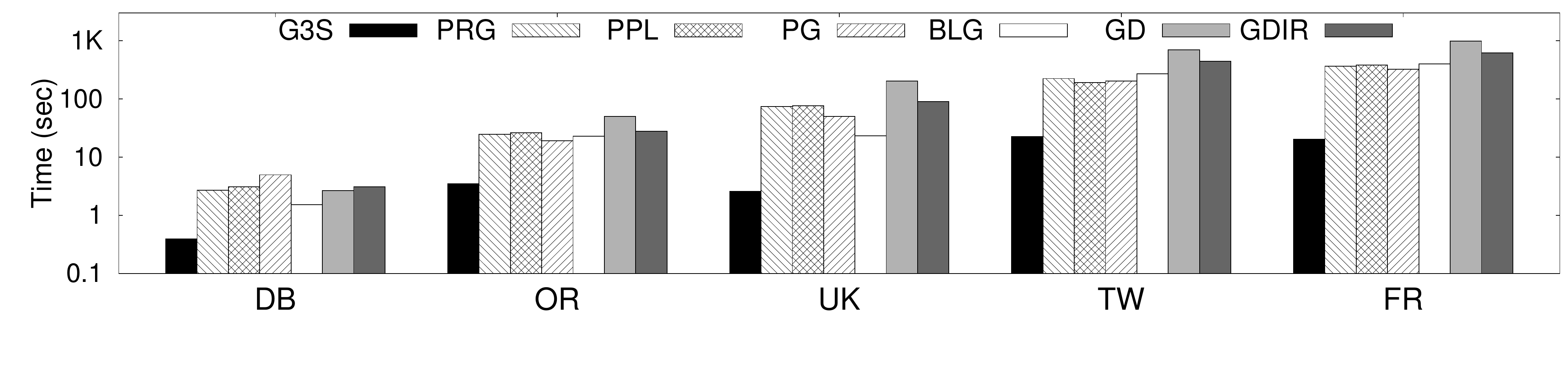}
	}
	\hfill
	\subfigure[\core]{ \label{fig:general_core}
		\includegraphics[width=0.485\linewidth, height=2.1cm]{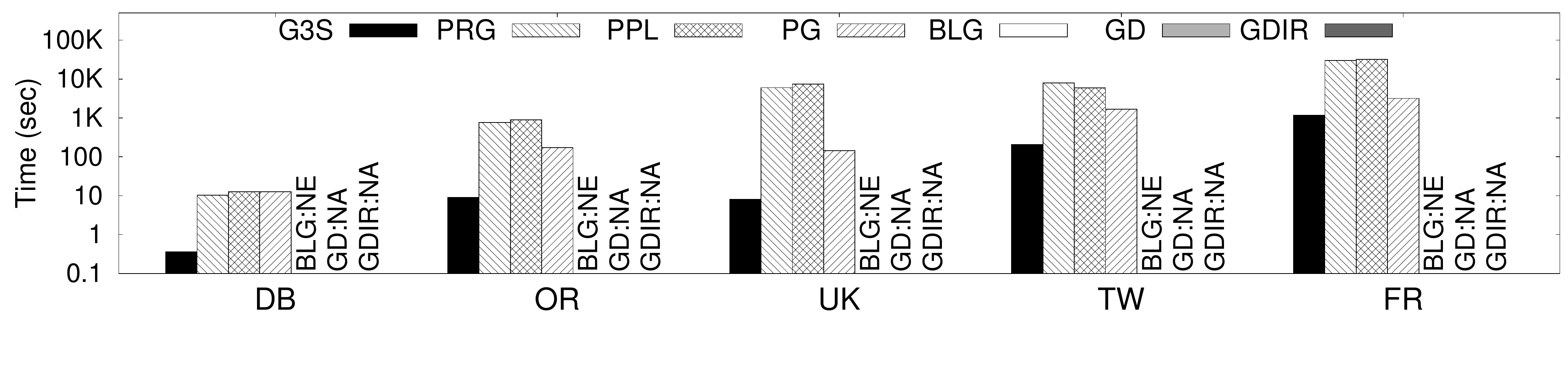}
	}
	\hfill
	\subfigure[\clr]{ \label{fig:general_color}
		\includegraphics[width=0.485\linewidth, height=2.1cm]{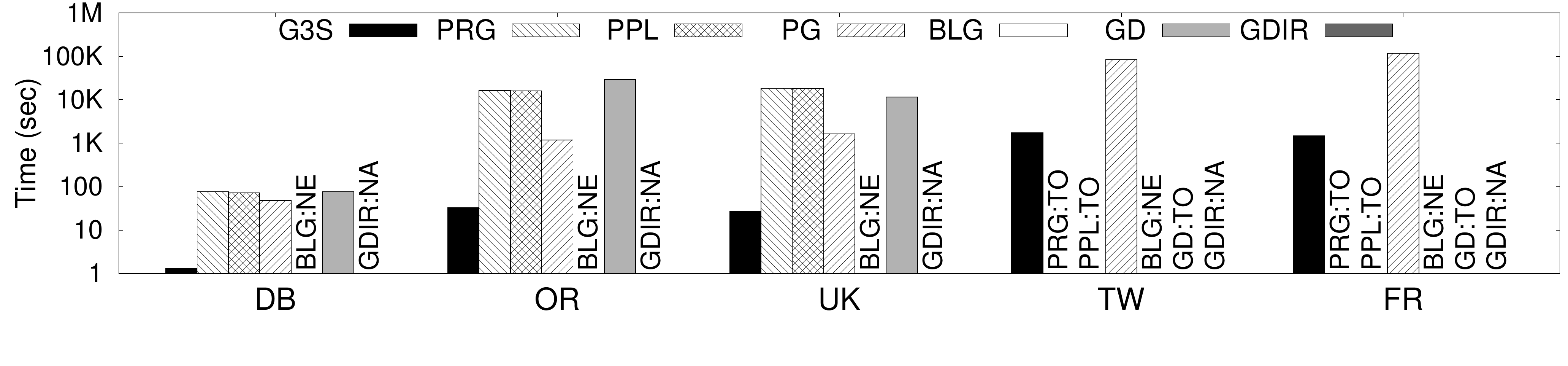}
	}
	\hfill
	\subfigure[\mis]{ \label{fig:general_mis}
		\includegraphics[width=0.485\linewidth, height=2.1cm]{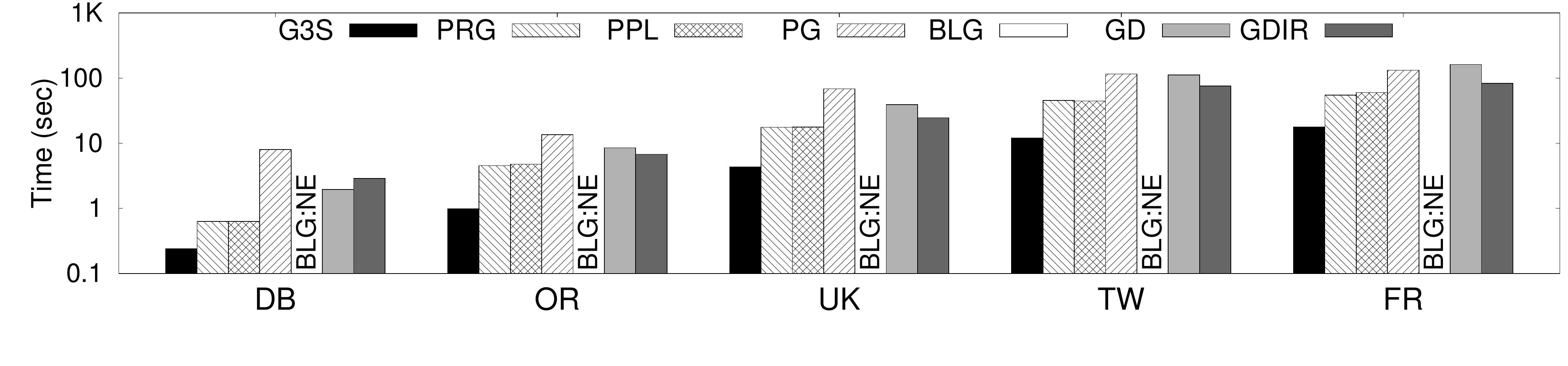}
	}
	\hfill
	\subfigure[\mm]{ \label{fig:general_mm}
		\includegraphics[width=0.485\linewidth, height=2.1cm]{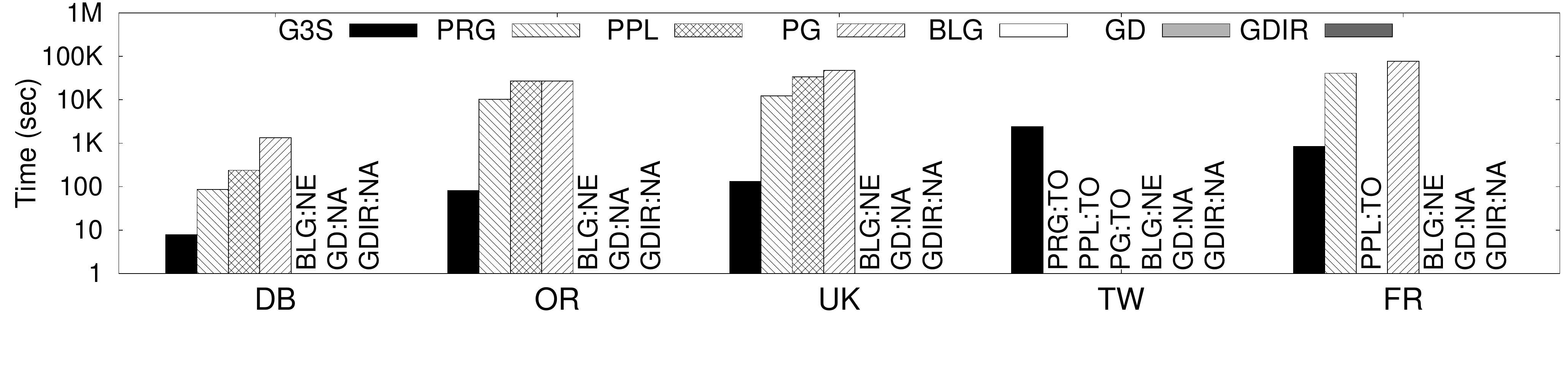}
	}
	\hfill
	\subfigure[\tric]{ \label{fig:general_tric}
		\includegraphics[width=0.485\linewidth, height=2.1cm]{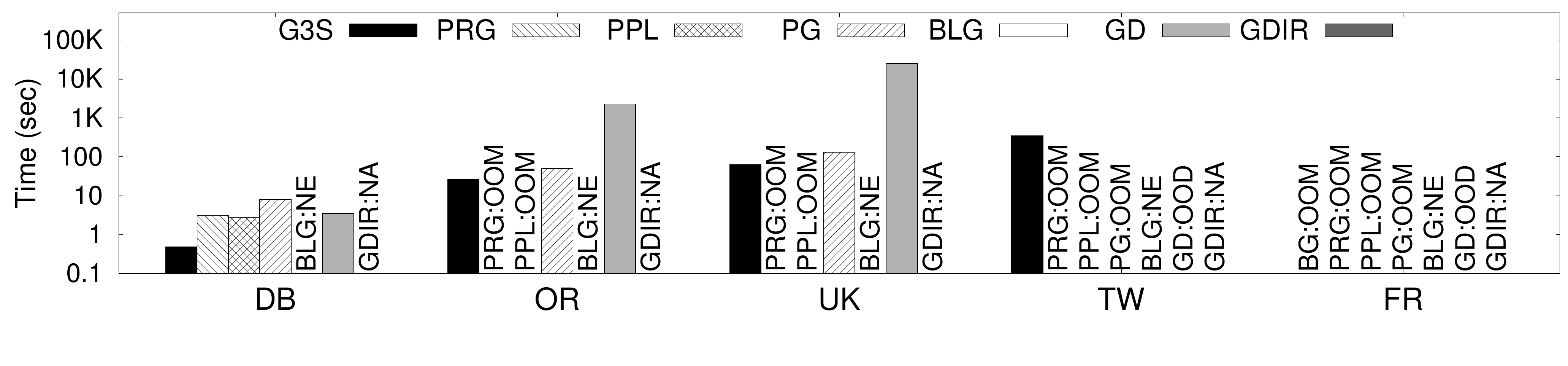}
	}
	\vspace{-0.15cm}
	\caption{Comparison with Existing Systems (Running Time)}
	\label{fig:general}
	%\vspace{-0.1cm}
\end{figure*}

\begin{figure*}[!h]
	\subfigure[\bfs]{ \label{fig:comm_bfs}
		\includegraphics[scale=0.2]{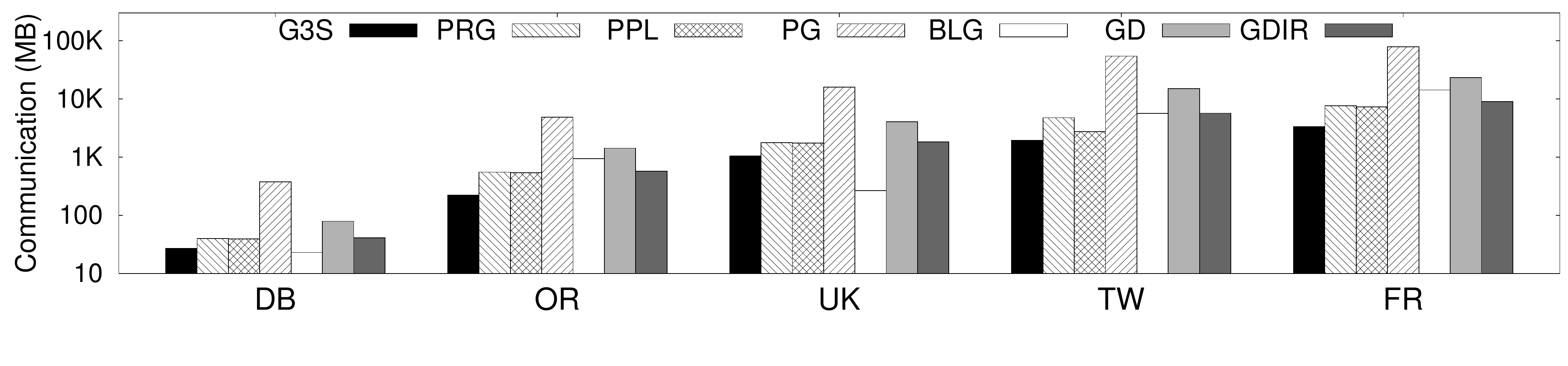}
	}
	\hfill
	\subfigure[\ccomp]{ \label{fig:comm_cc}
		\includegraphics[scale=0.2]{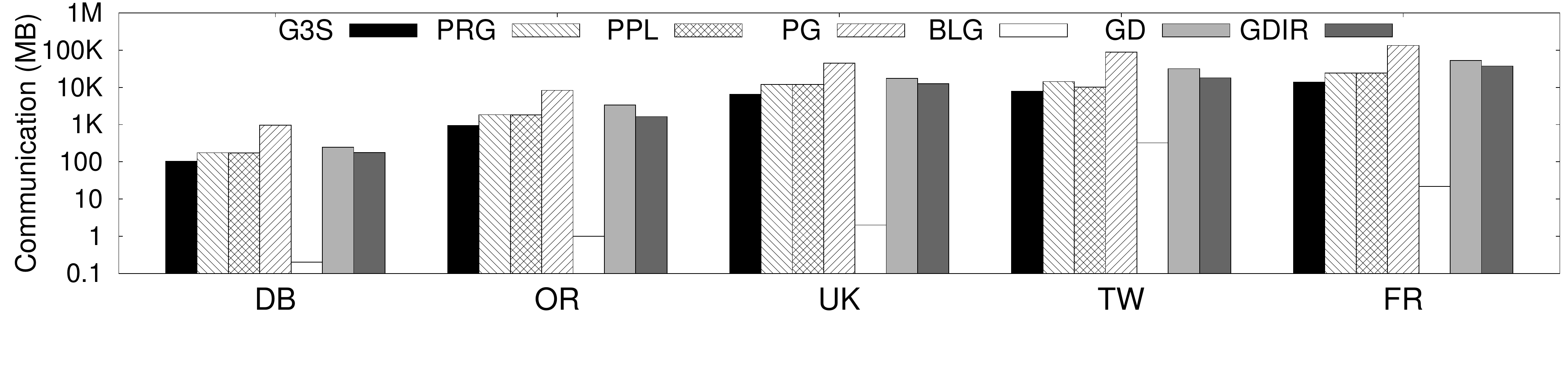}
	}
	\hfill
	\subfigure[\pr]{ \label{fig:comm_pr}
		\includegraphics[scale=0.2]{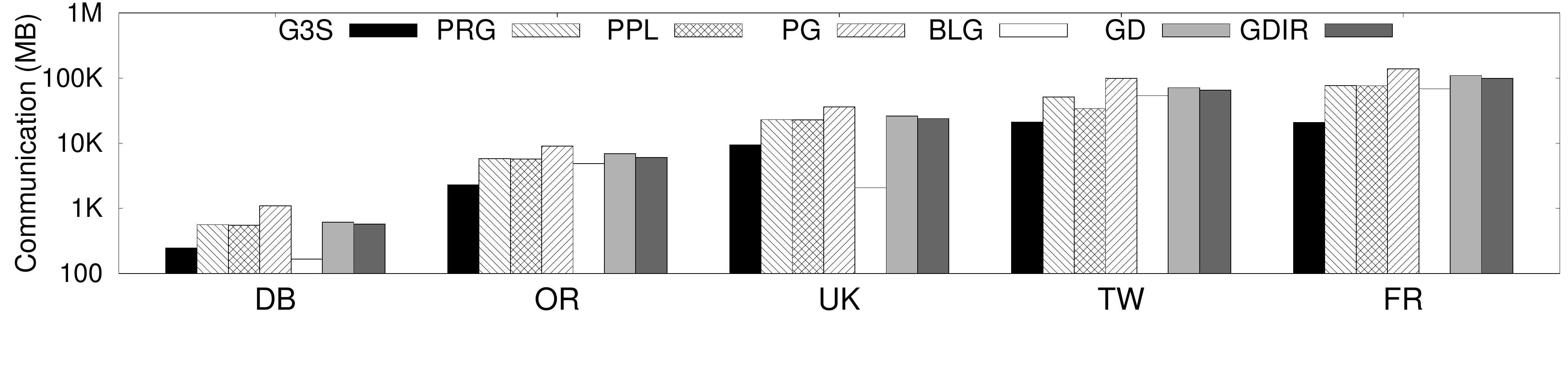}
	}
	\hfill
	\subfigure[\ppr]{ \label{fig:comm_ppr}
		\includegraphics[scale=0.2]{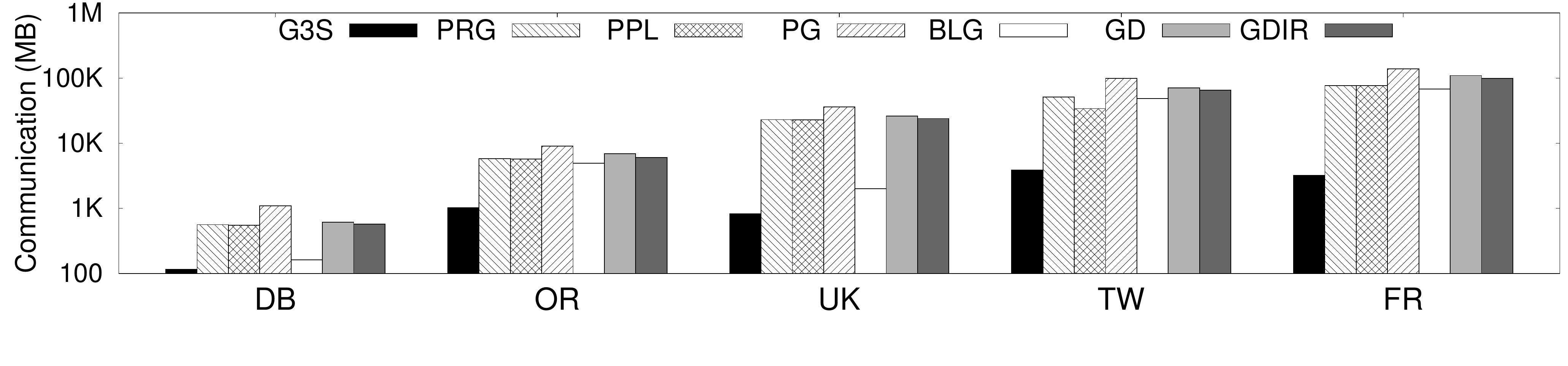}
	}
	\hfill
	\subfigure[\core]{ \label{fig:comm_core}
		\includegraphics[scale=0.2]{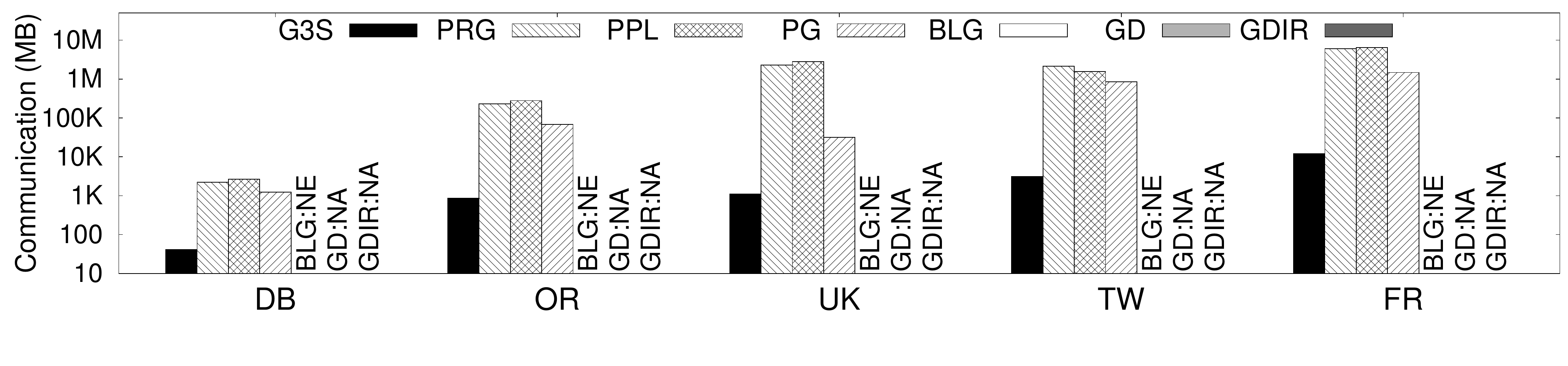}
	}
	\hfill
	\subfigure[\clr]{ \label{fig:comm_color}
		\includegraphics[scale=0.2]{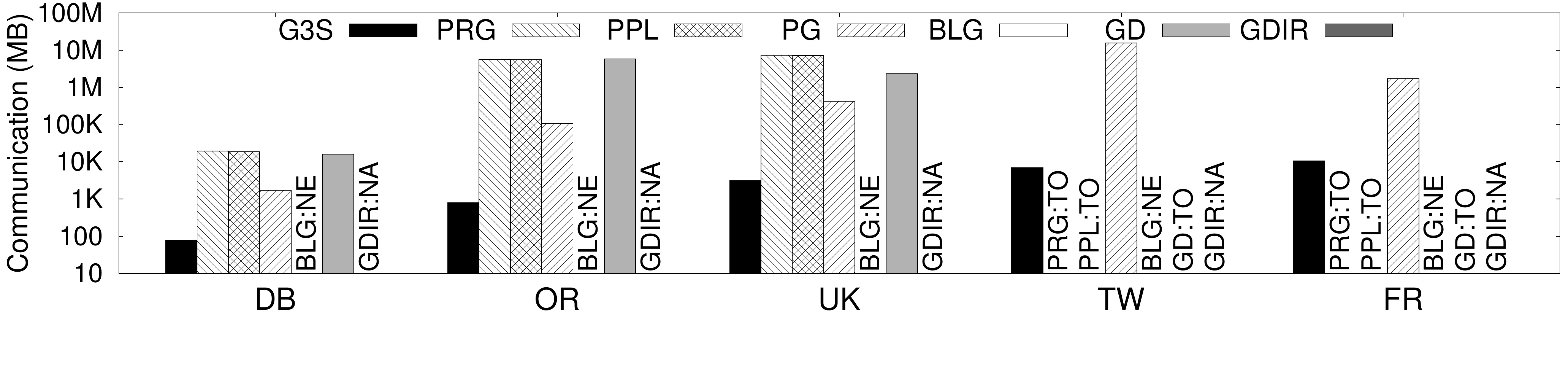}
	}
	\hfill
	\subfigure[\mis]{ \label{fig:comm_mis}
		\includegraphics[scale=0.2]{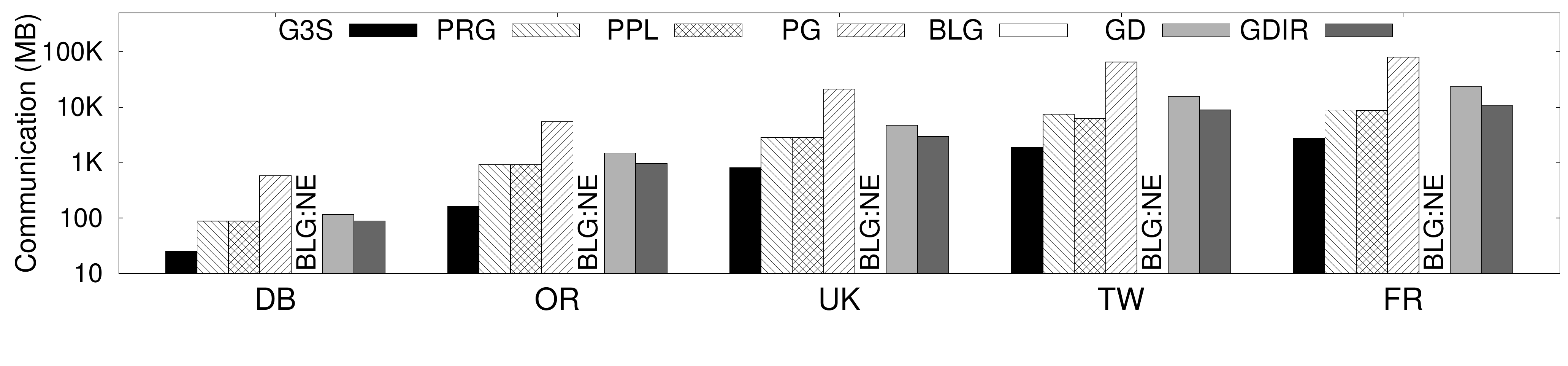}
	}
	\hfill
	\subfigure[\mm]{ \label{fig:comm_mm}
		\includegraphics[scale=0.2]{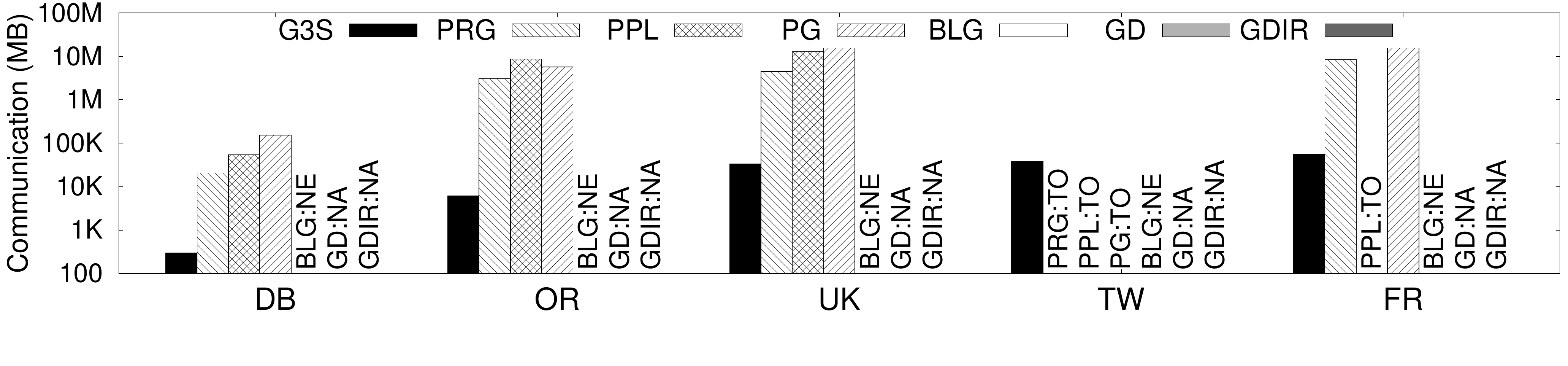}
	}
	\hfill
	\subfigure[\tric]{ \label{fig:comm_tric}
		\includegraphics[scale=0.2]{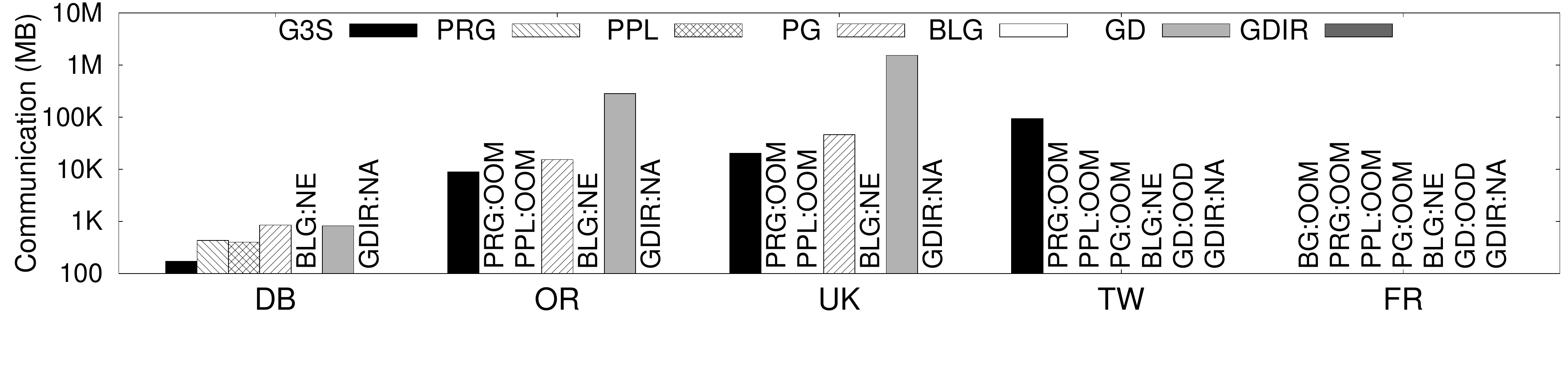}
	}
	\vspace{-0.15cm}
	\caption{Comparison with Existing Systems (Communication Cost)}
	\label{fig:comm}
	%\vspace{-0.1cm}
\end{figure*}

\subsection{Efficiency over Different Algorithms}
\label{sec:exp2}

We compared the system efficiency when running different algorithms over given datasets. The running time results are shown in \cfig \ref{fig:general}. 
We use NE and NA to represent Not Effective and Not Applicable cases as mentioned above respectively. OOM, OOD and TO represent Out Of Memory, Out Of Disk and Time Out respectively. We consider an algorithm running as time out when it can't finish within 24 hours. 

\begin{figure*}[!h]
	\subfigure[{\small \bfs (\tws) }]{ \label{fig:scalet_bfs}
		%\centering
		\includegraphics[scale=0.28]{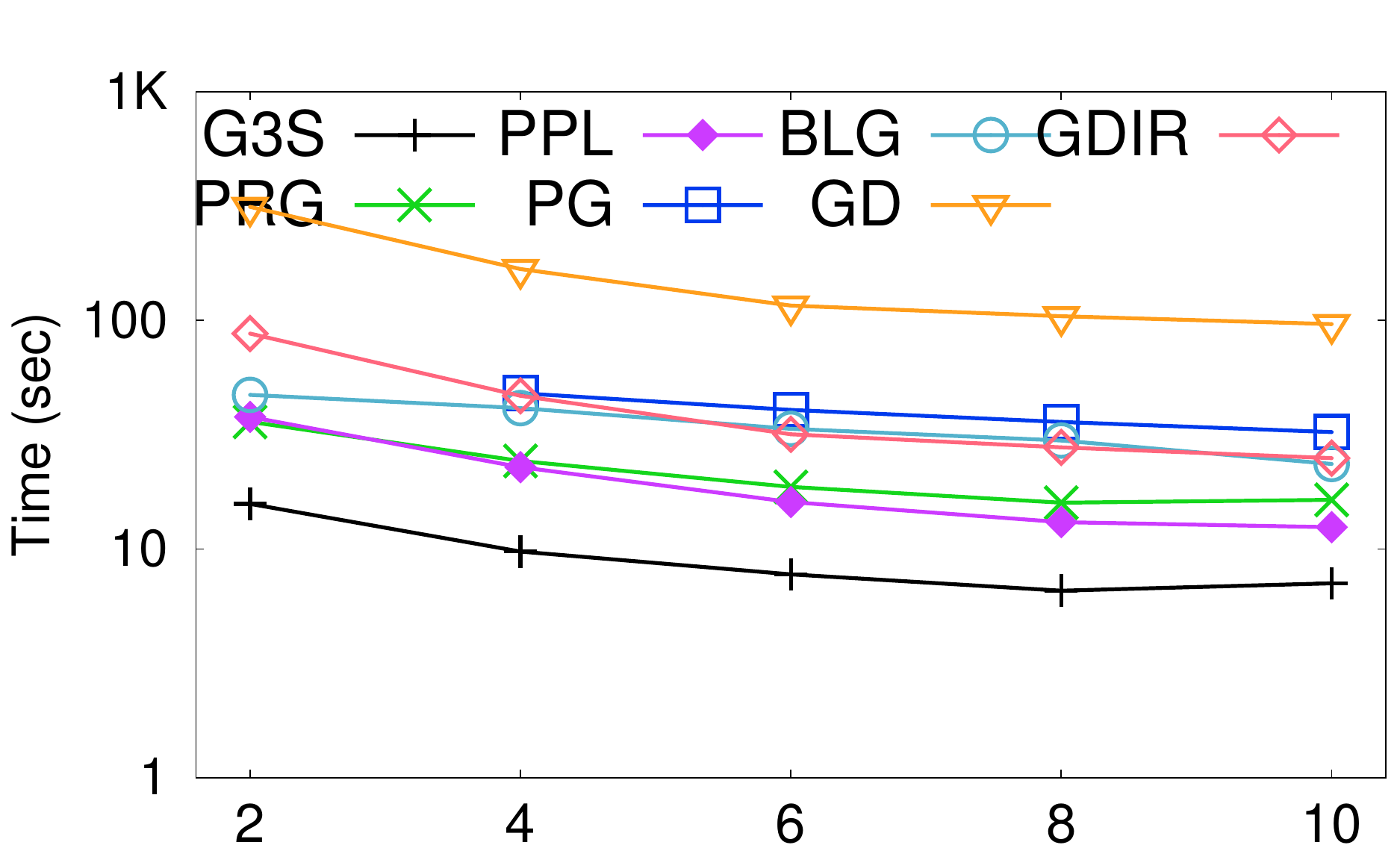}
	}
	\hfill
	\subfigure[{\small \pr (\tws)}]{ \label{fig:scalet_pr}
		%\centering
		\includegraphics[scale=0.28]{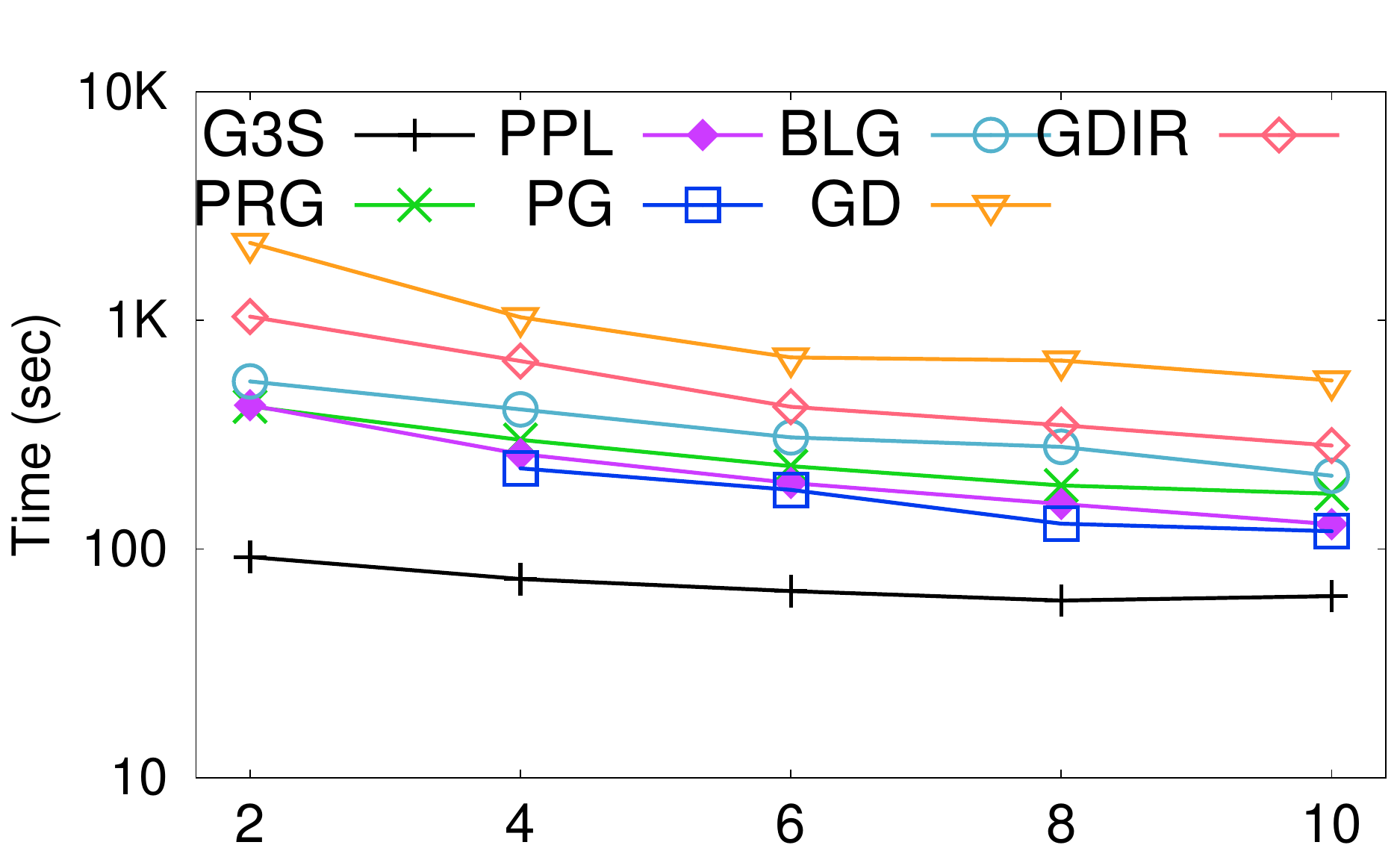}
	}
	\hfill
	\subfigure[{\small \core (\tws)}]{ \label{fig:scalet_core}
		%\centering
		\includegraphics[scale=0.28]{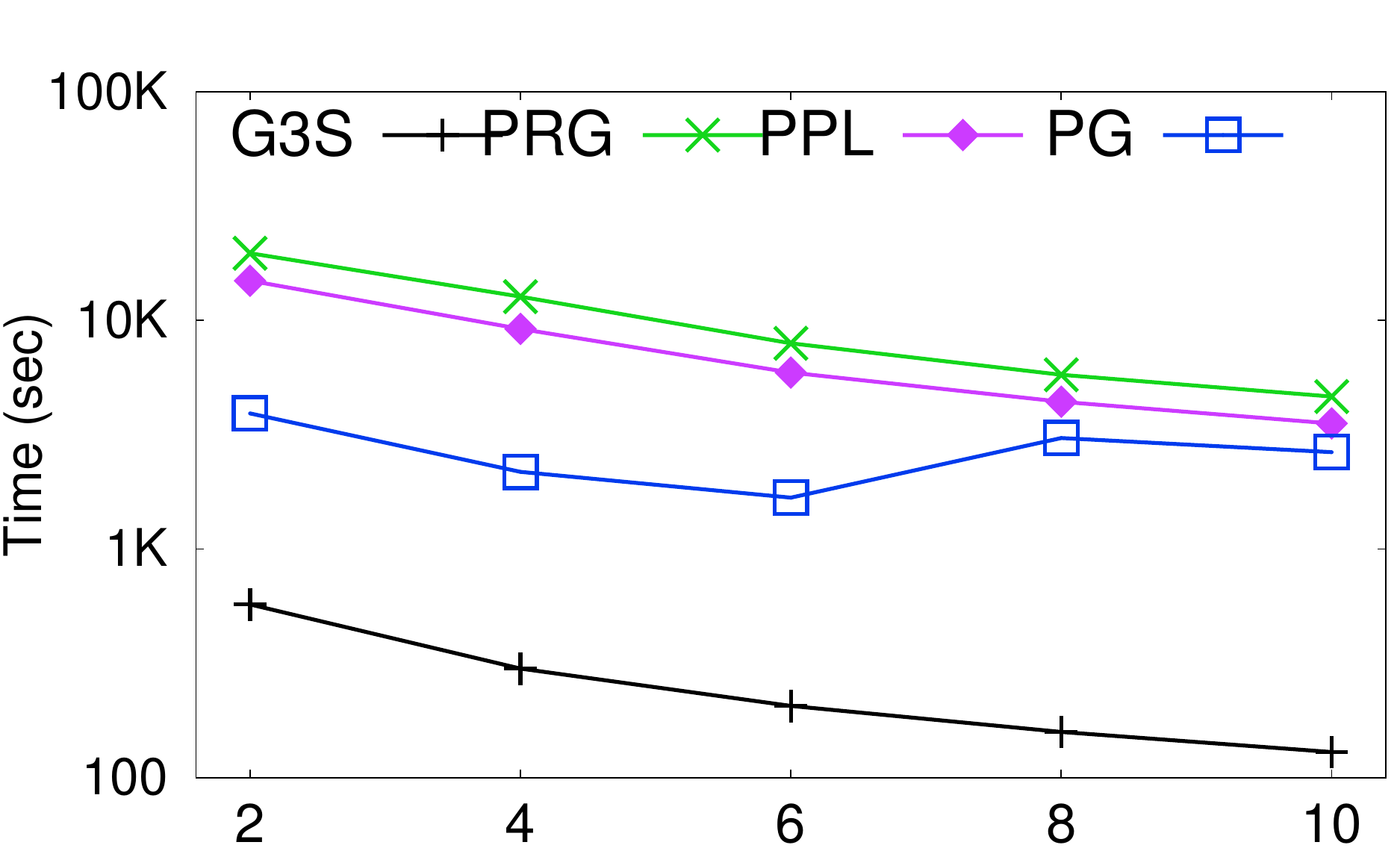}
	}
	\hfill
	\subfigure[{\small \mm (\tws)}]{ \label{fig:scalet_mm}
		%\centering
		\includegraphics[scale=0.28]{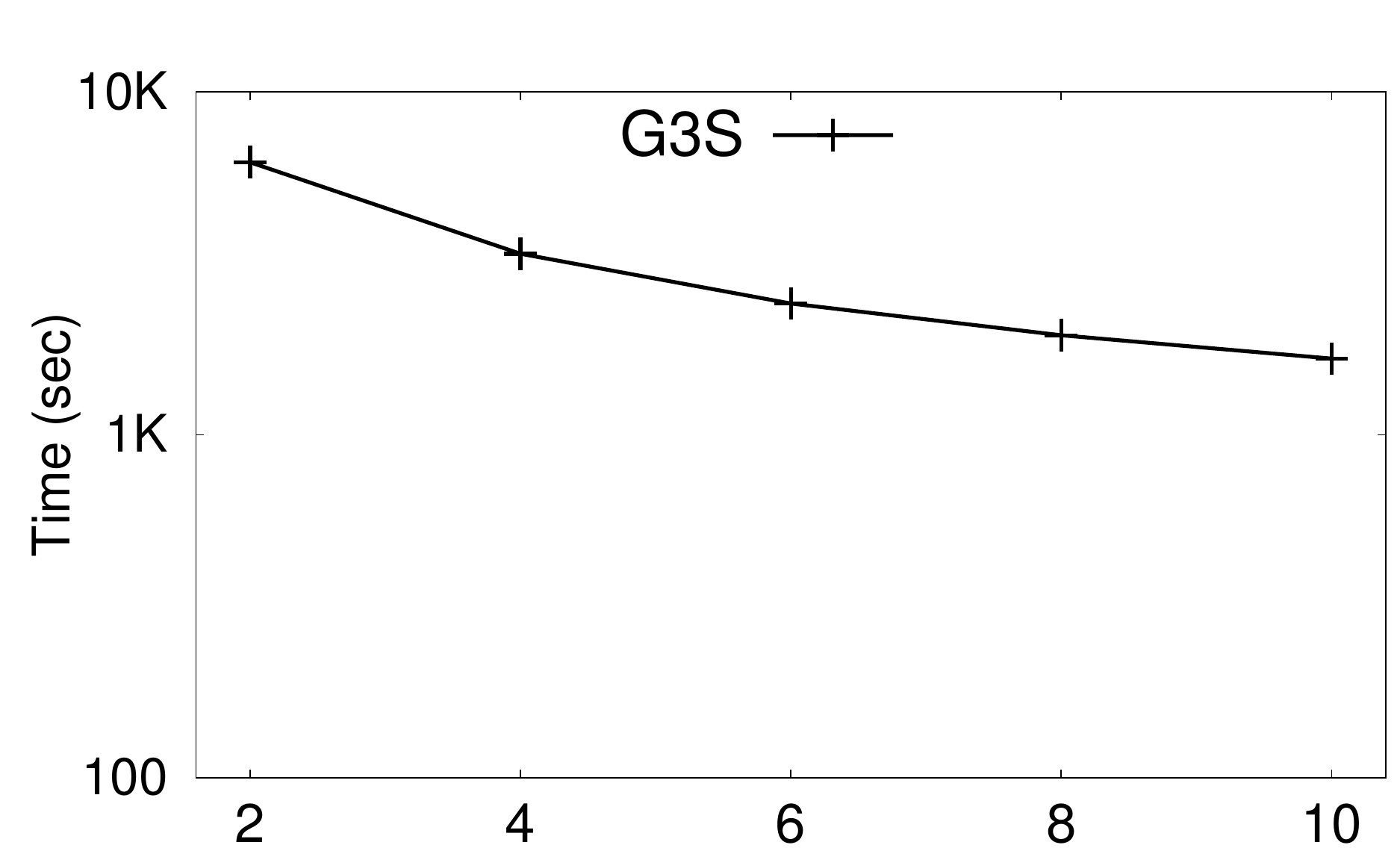}
	}
	\hfill
	\subfigure[{\small \tric (\tws)}]{ \label{fig:scalet_tric}
		%\centering
		\includegraphics[scale=0.28]{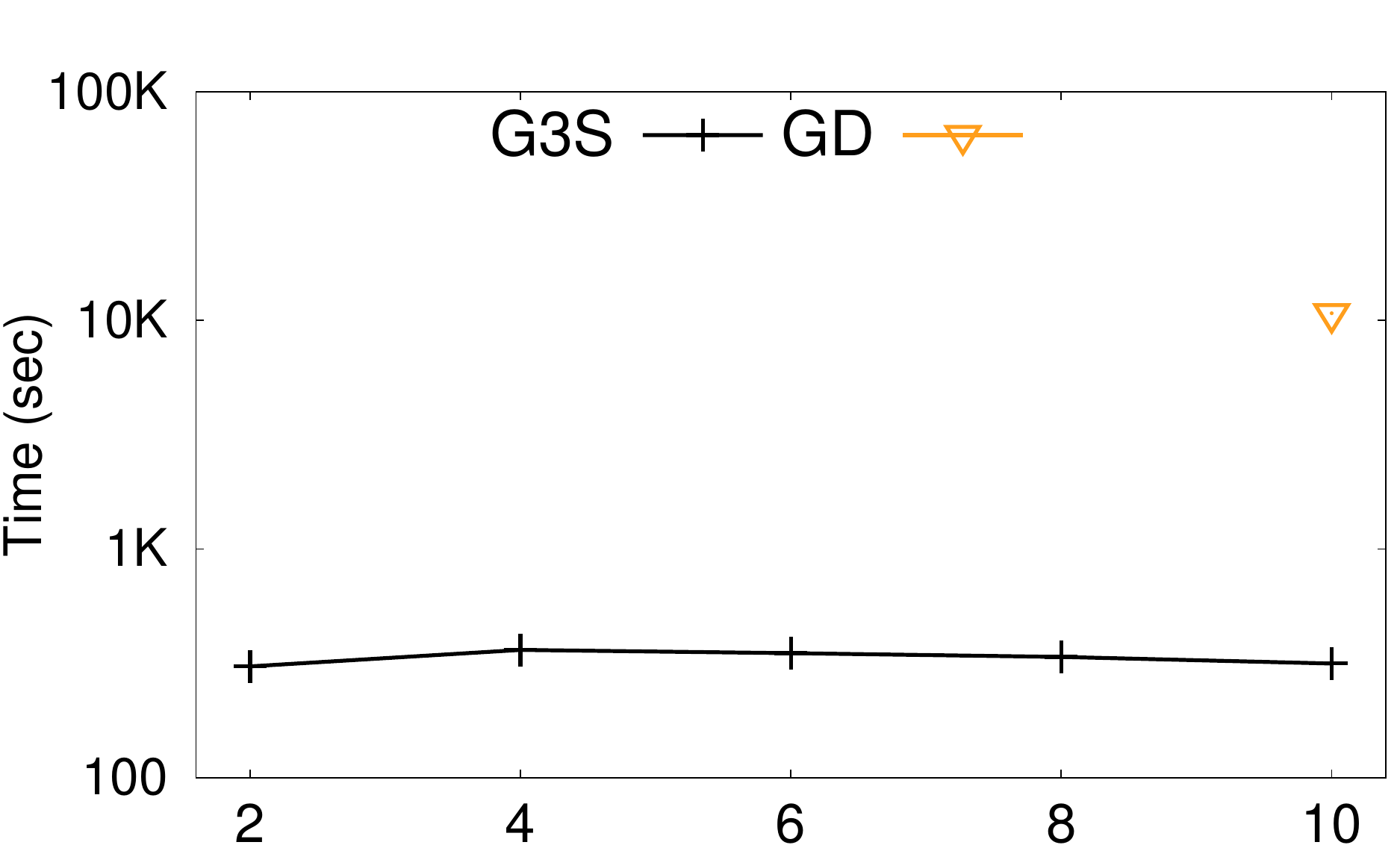}
	}
	\hfill
	\subfigure[{\small \bfs (\frs)}]{ \label{fig:scalef_bfs}
		%\centering
		\includegraphics[scale=0.28]{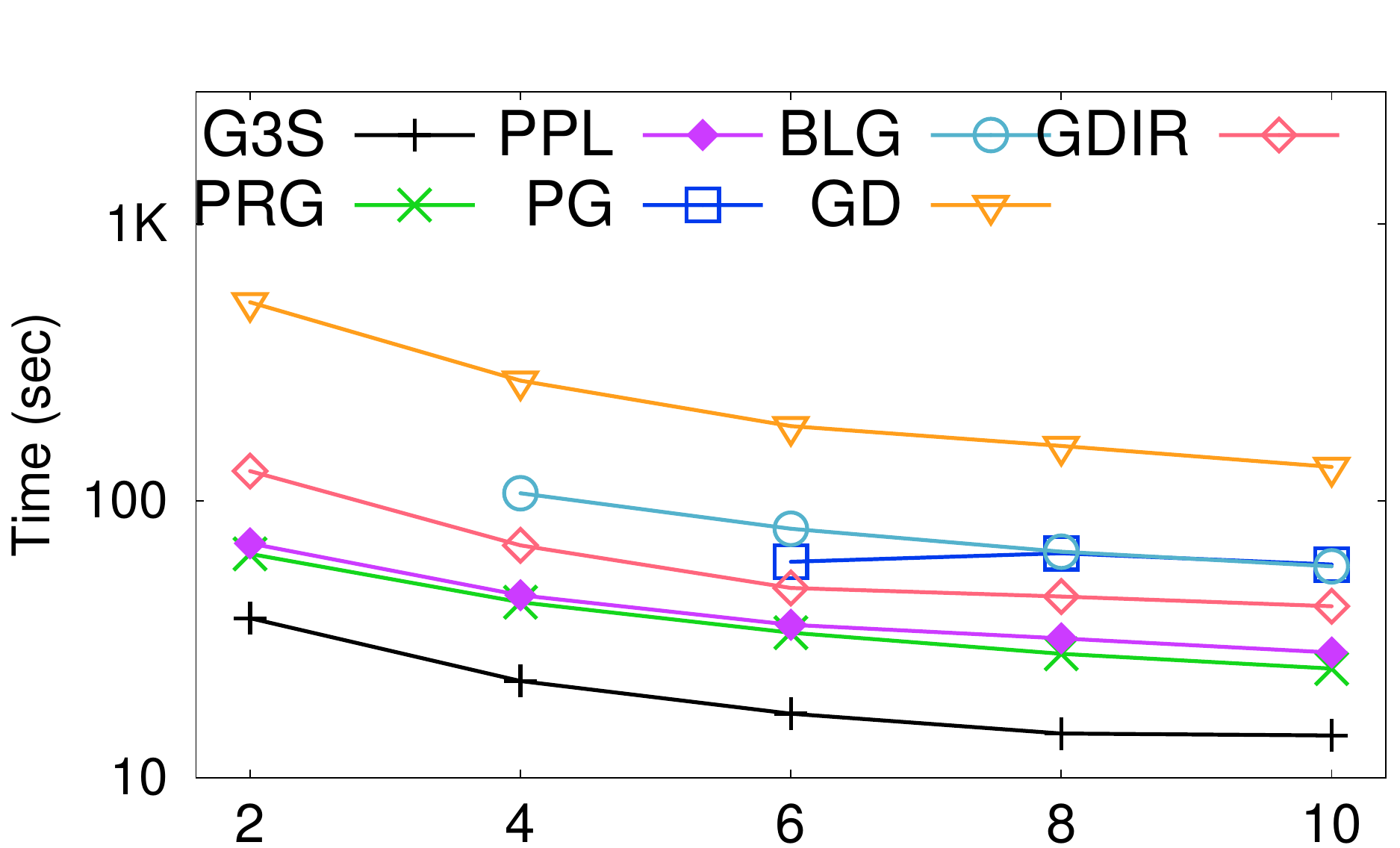}
	}
	\hfill
	\subfigure[{\small \pr (\frs)}]{ \label{fig:scalef_pr}
		%\centering
		\includegraphics[scale=0.28]{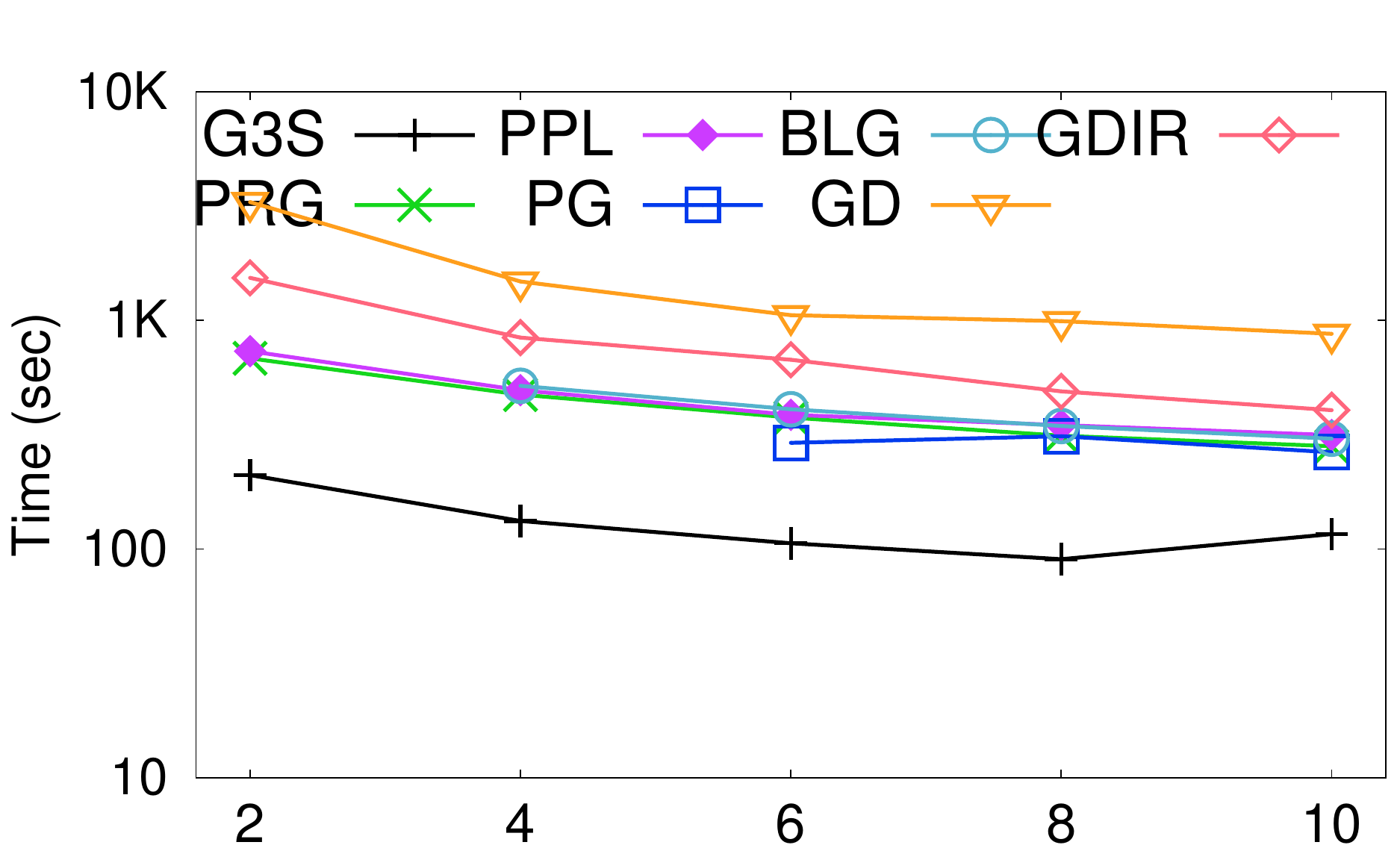}
	}
	\hfill
	\subfigure[{\small \core (\frs)}]{ \label{fig:scalef_core}
		%\centering
		\includegraphics[scale=0.28]{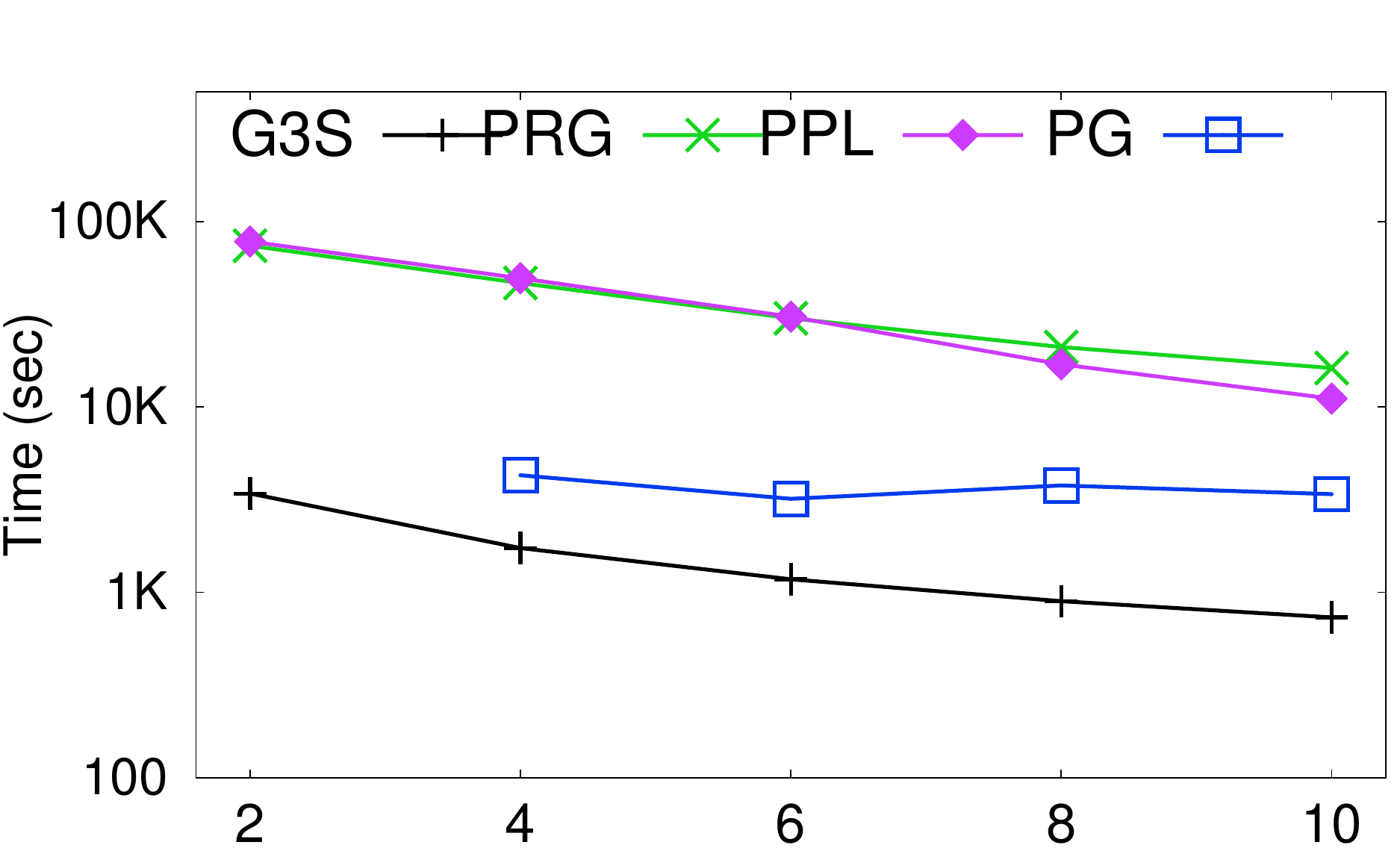}
	}
	\hfill
	\subfigure[{\small \mm (\frs)}]{ \label{fig:scalef_mm}
		\centering
		\includegraphics[scale=0.28]{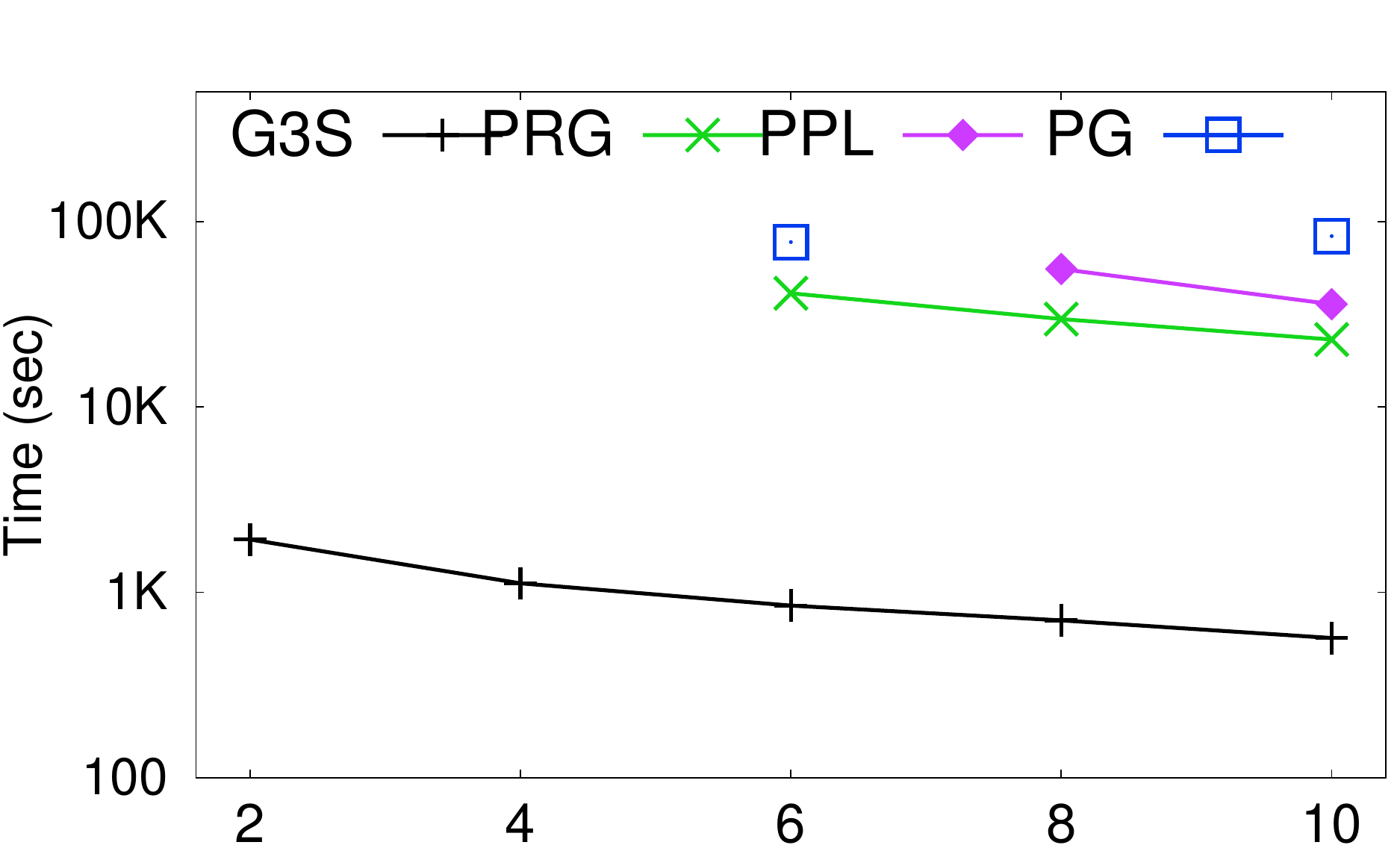}
	}
	%\vspace{-0.1cm}
	\caption{Scalability Test (Varying \#machines)}
	\label{fig:scale}
\end{figure*}

From the results, we can see that, except for \ccomp on \blgs, \newg outperforms all compared systems. 
For separable algorithms including \bfs, \ccomp, \pr, \ppr and \mis, \newg outperforms \prgs, \ppls, \pgs, \blgs, \gd and \gdir by 5.2x, 5.3x, 9.8x, 4.1x, 14.7x and 7.8x on average respectively. These algorithms are commonly used for system performance evaluations in existing works. \newg shows good speedup results. 
This proves the effectiveness of our speedy techniques as proposed in \csec \ref{speedy}. 
Without the trouble of considering and designing combiners as with existing systems, the critical attributes, dual neighbor index and self-adaptive activation of \newg are able to guarantee a strong performance by saving computation and communication cost.
For non-separable algorithms including \core, \clr, \tric and \mm, many optimizations in existing works like combiners, ID recoding are not applicable. Thus the outperformance of \newg is more severe in these cases. For example, \newg outperforms \prgs, \ppls, \pgs, and \gd by 198.2x, 249.3x, 89.4x, and 311.8x on average respectively. The speedup can even reach 906.0x when running \core on \uk compared to \ppls. Note that the cases that systems can't finish running algorithms are not included. These results demonstrate the consistent outperformance of our system over diverse algorithms. 
%
%
%
% pregel & pregelplus
Among existing systems, vertex-centric systems have similar performances. But for separable algorithms like \bfs, \ccomp and \mis, \prgs and \ppls performs better than \pgs because of the combiner used effectively saves communication cost. One exception is \pr and \ppr because \pgs adopts delta caching. For non-separable algorithms like \core, \clr and \mm, a combiner is inapplicable leaving the superiority of \prgs and \ppls lacking.
In terms of \prgs and \ppls, although \ppls has a mirroring technique, it cannot always beat \prg. This is because the cost saved by the mirroring technique does not always compensate for the additional cost of transferring mirror node messages.
Another problem with \ppls is that the mirror threshold needs to be designated by users which makes it difficult to achieve best system performance. This requires a user to be familiar with the system, algorithm and used dataset. 
A lower threshold will cause higher communication costs and memory consumption. At a certain point, it may cause OOM failure. This is why the authors recommend the threshold be at least 100 for large graphs. However, a too large threshold reduces the cost saving by mirroring technique.
In fact, even the cost model given in their paper does not guarantee the best performance. 
% Besides, the performance of \ppls depends on mirror threshold. 
% combiner
%Also, combiner is an important factor affecting the communication cost of \prg and \ppl. For non-separable algorithms like \core, \clr and \mm, they both have high communication cost which affects its total performance. 
% powergraph
%\pg outperforms \prg and \ppl over most datasets running \pr, \ppr, \core and \clr. This is because 
% blogel
The block-centric system \blgs performs best among all compared systems running \ccomp. This is because the pre-processing partition of \blgs already partitions connected subgraphs into the same machine. It favors the algorithm running within connected subgraphs because the vertices inside a block can keep computing until no more updates apply. In this way, the total number of algorithm supersteps is reduced. %Similarly, \blgs is able to outperform \newg in terms of \bfs on \uk and is competitive in terms of \pr and \ppr. 
% GraphD
The performance of \gd and \gdir are consistent with results from \cite{yan2018graphd}. The results also show the effectiveness of IR recoding technique on separable algorithms. Generally, \graphd can't beat in-memory vertex-centric systems. This is because it focuses on system scalability and is involved with disk accesses. Its better performance  will be shown in the following scalability tests.

%\subsection{Communication Cost Comparision}
\stitle{{Communication Cost Comparison}}
We also report the communication cost comparison results of evaluated systems in \cfig \ref{fig:comm}.
The results are mostly consistent with the running time presented above. More communication cost leads to longer running time. 
In most cases, \newg generates the least communication. %except \blgs on \bfs and \ccomp. The reason is discussed above. 
%For example, 
This benefits from our critical attributes design which saves unnecessary attributes transformation cost. 
\prgs and \ppls have similar communication costs. %Which one is better depends on whether 
\pgs incurs the highest communication cost in most tests. This is because other systems have different techniques to reduce communication cost, like combiner design and mirroring technique.
\blgs has less communication cost in \pr and \ppr because we only report the communication cost of B-mode (same with running time). The whole program needs to run V-mode first. 
Without ID recoding, \gd has more communication cost than \gdir and \gdir generates similar cost with \prgs and \ppls.

\subsection{Scalability Test}
\label{sec:exp3}

In this section, we evaluate the scalability of all systems by varying the number of tested graph size  and used machines respectively. Due to the space limit, we choose representative algorithms \bfs, \pr, \core, \mm and \tric to report the results in this part.

\stitle{Varying the Number of Machines.}  Firstly, we test the scalability of \newg in comparison with existing systems by varying the number of machines. For each machine, all four cores are used. We run selected algorithms over two large datasets \twit and \friend. The results are shown in \cfig \ref{fig:scale}. Note that the result of running \tric on \friend is not shown because only \gd manages to finish when 10 machines are used within 24 hours (3800.24s).

The experimental results show that, in most cases, with the increasing number of machines used, better efficiencies are achieved. 
This is because the greater number of machines used, more parallel computation happens and the less computation time consumes. As a result, total running time reduces. 
However, more machines also means more communication cost. So, when saved computation cost doesn't compensate increased communication cost, the total time couldn't be reduced but will increase. This explains in some cases how the more machines used, more time is consumed. For example, running \core on \pgs over \twit, the total time increases when eight machines used compared to six machines.

We also find that the existing systems have different favourable algorithms. 
%Vertex-centric systems are strong in efficiency and disk-based systems are good in scalability. 
%
%Vertex-centric systems performs well for cpu-intensive algorithms like \bfs, \pr and \core (still outperformed by \newg) but lose advantages in memory-intensive algorithm like \tric. For example, they can't finish in 24 hours running \tric on either \twit or \friend.  
%%
%On the contrary, disk-based system \graphd shows poor performance for cpu-intensive algorithms but performs well for memory-intensive algorithms. For example, \graphd is the only system which finishes running \tric on \friend. This is understandable because \graphd uses disk to increase system scalability which meanwhile increases both computation and communication cost. 
%
For example, in-memory systems show better performance running cpu-intensive algorithms like \bfs, \pr and \core. While out-of-core system \graphd shows poor efficiency performance. For example, it can't finish in 24 hours running \core on either \twit or \friend. This is understandable because \graphd uses disk to increase system scalability which meanwhile increases both computation and communication costs.
However, for memory-intensive algorithms like \tric,  disk-based system \graphd shows better performance than in-memory systems that usually run out of memory. For example, \graphd is the only system that finishes running \tric on \friend.

Different from existing systems, \newg shows excellent overall performance for all kinds of algorithms. 
In terms of cpu-intensive algorithms, it outperforms existing in-memory systems. For example, it is averagely 36.6, 27.1 and 11.4 times faster than \prgs, \ppls and \pgs respectively running \core on \twit for different number of machines.  
%Note that \newg is the only system that finishes running \mm over \twit within 24 hours even when only 2 machines are used. 
%
For memory-intensive algorithms, \newg is competitive compared to \gd. For instance, \newg and \gd are the only two systems that finish running \tric on \twit. 
It is worth noticing that \gd can only finish when all ten machines are used. Nevertheless, \newg is able to finish even when only two machines are used. Adding to this, though both systems finish when all 10 machines are used, \newg is 33.1 times faster than \gd. 
This further demonstrates the good combinational performance of \newg compared with existing systems.

%Among the compared systems, the vertex-centric systems like \prg, \ppl and \pg have compatative performances over the others in terms of efficiency. 
%%
%However, for memory-intensive algorithms like \tric, they all run out of memory on \twit and \friend. While disk-based system \graphd can get result when all 10 machines are used on \twit. Note that \graphd is the only system, excluding \newg, that finishes \tric on \friend. This is because \graphd streams edge and message data on local disks and thus consumes neligible memory space.
%%
%Note that, as we mentioned earlier, \core and \mm are not implemented on \graphd because the system desiganation of \graphd does not support.

\begin{figure}[t]
	\subfigure[{\small \bfs }]{ \label{fig:scalee_bfs}
		%\centering
		\includegraphics[scale=0.205]{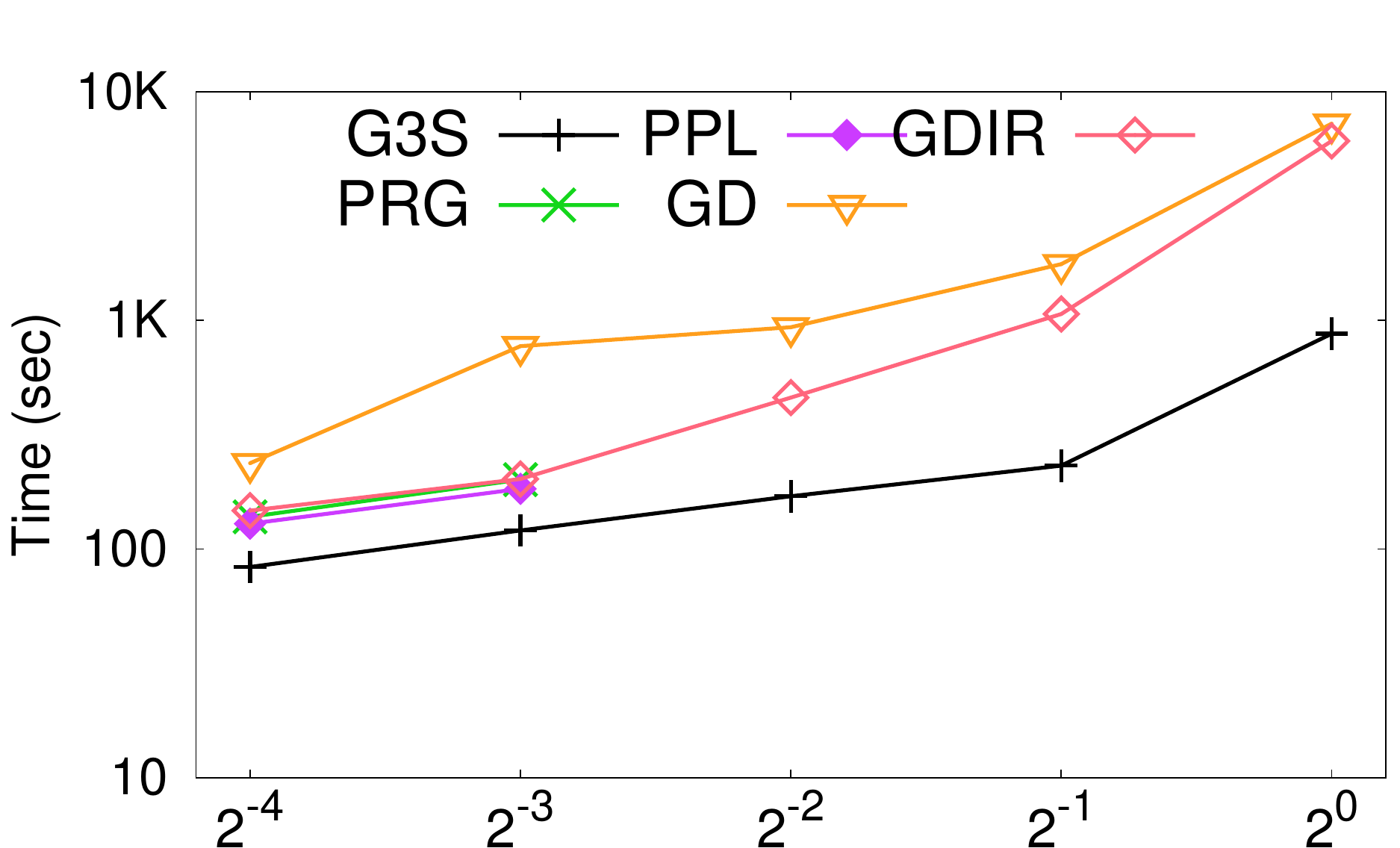}
	}
	\hfill
	\subfigure[{\small \pr}]{ \label{fig:scalee_pr}
		%\centering
		\includegraphics[scale=0.205]{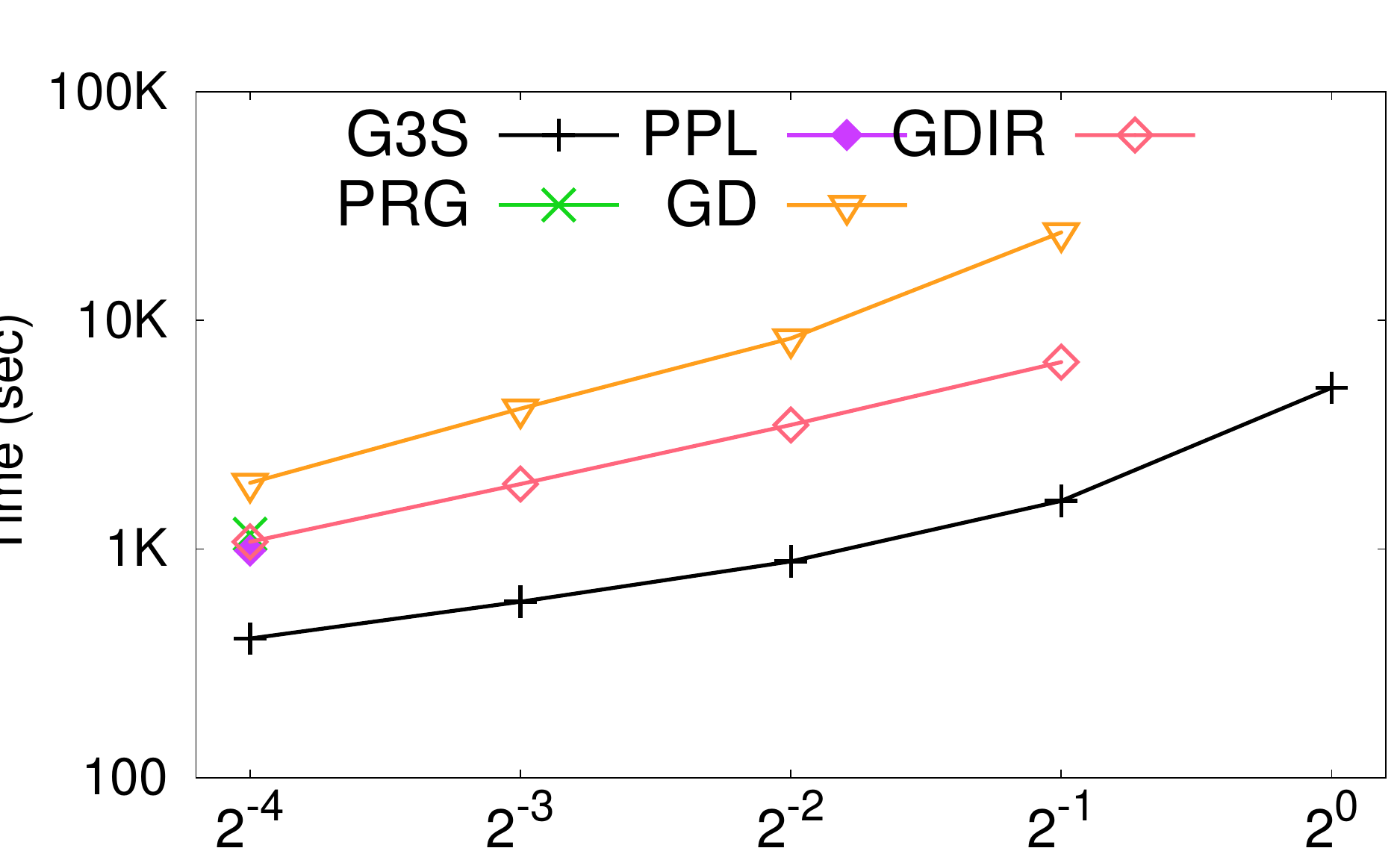}
	}
	\hfill
	\subfigure[{\small \core}]{ \label{fig:scalee_core}
		%\centering
		\includegraphics[scale=0.205]{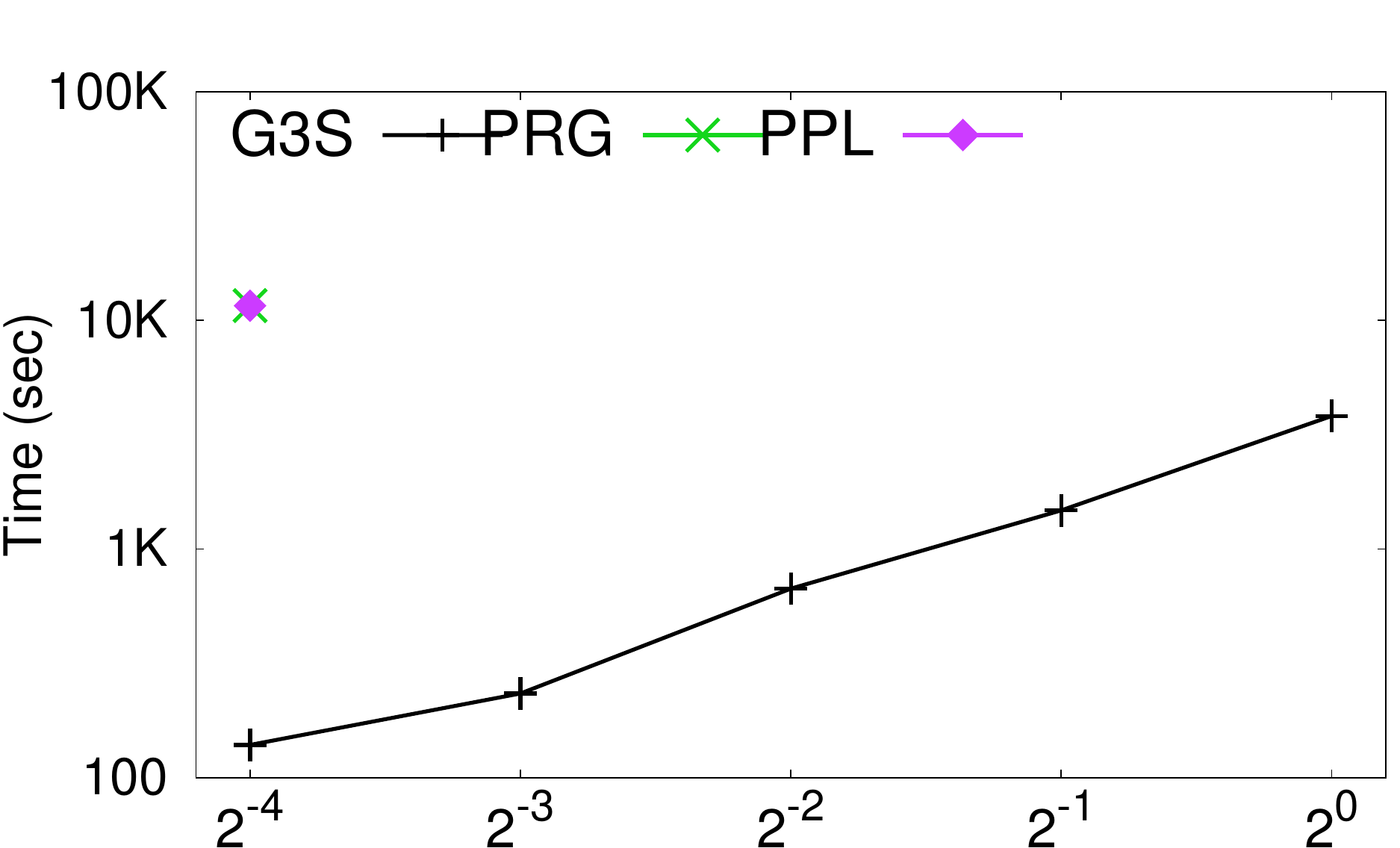}
	}
	\hfill
	\subfigure[{\small \mm}]{ \label{fig:scalee_mm}
		%\centering
		\includegraphics[scale=0.205]{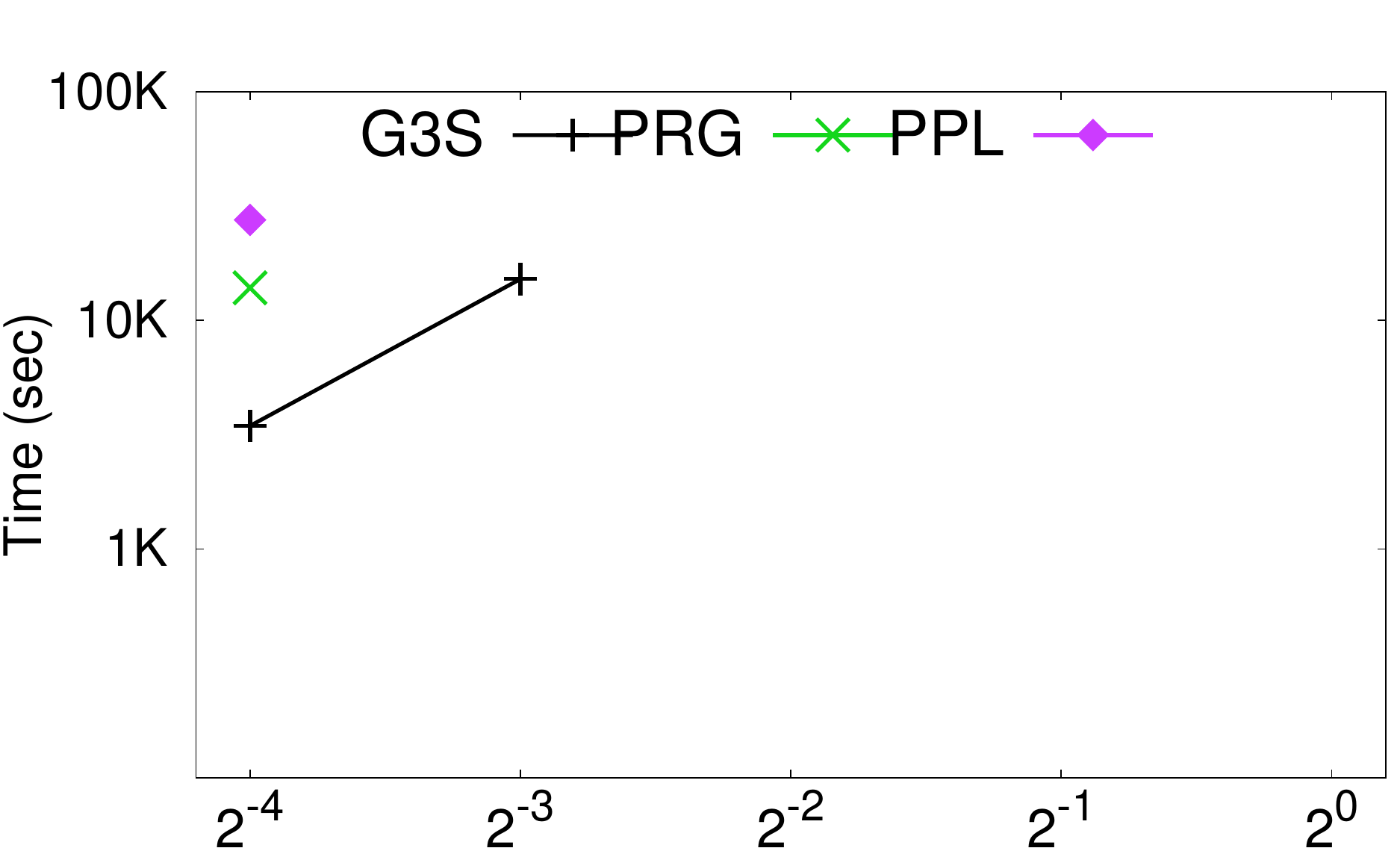}
	}
	\caption{Scalability Test (Varying \#edges)}
	\label{fig:scalee}
\end{figure}

\stitle{Varying Graph Size.}
We also test the system scalability by varying graph size. We adopt the largest used dataset \clueweb and sample $2^{-1}$, $2^{-2}$, $2^{-3}$, $2^{-4}$ of all edges to vary the graph size. The experimental results are shown in \cfig \ref{fig:scalee}. Note that \tric results are not reported because no system can finish running \tric on any used dataset within 24 hours. Also, \pgs and \blgs are not shown because they both couldn't finish on used big graphs.

The results show that with dataset size increasing, the running time of all systems increases as well. All systems show similar increasing behavior. The results are consistent with that in \csec \ref{sec:exp2}. Our new system \newg shows superb scalability performance. \prgs and \ppls show good efficiency but can't finish when the graph is too large. 
Disk-based system \graphd shows good scalability but is weak in efficiency. Especially for non-separable algorithms where ID recoding (\gdir) is inapplicable, \gd is very time-consuming. 
%For example, 
For example, when all edges in \clueweb are used in \pr, \newg is the only system that can finish within 24 hours.

\subsection{Other Issues}

%\newg can only be applied in graph applications where vertex only communicates with neighbor vertices. 

Fault tolerance is important to a system. It is not considered in current work because the authors of \pg state in their paper \cite{gonzalez2012powergraph} that the overhead, typically a few seconds for largest graph used, is relatively small compared to the total running time. This is consistent with our experiments. For example, the total time of \pgs for running \clr on \twit, a dataset also used in their paper, is 23 hours. However, we leave the implementation of \newg's fault tolerance in the future. 
Also, asynchronous mode is not considered because it is not general and is only effective on algorithms with asymmetric convergence behavior and low workload \cite{yan2017big}. 

%\vspace*{-0.2cm}
\section{Conclusion}
\label{sec:conclusion}

%This paper studies usage simplicity of distributed graph processing systems which has not been well discussed in the literature. A simple programming model \newm is designed where users can think like a neighborhood expression and a system \newg is built to implement the model. 
%
%To the best of our knowledge, this is the first work which studies the usage simplicity of distributed graph processing systems.
%
%works system simply usage and also propose a simple, speedy and scalable graph processing system named \newg so that a user can design an application by just giving a neighborhood expression. \newg is very simple to use but also have good performance in efficiency and scalability compared to existing systems. We show the good performance by conducting extensive experiments over large graphs. 

%This paper contributes to the study of distributed graph processing systems by 
The main goal of this paper was to develop a system for a good combinational performance of all simplicity, efficiency and scalability. 
We provide an idea of achiveing the goal by trading vertex comuunication flexibility. A simple, speedy and scalable \newg is designed with a simple programming model and 
different optimization techniques guaranteeing its efficiency and scalability. 
Extensive experimental results demonstrate the outstanding performance of \newg compared to existing systems. 
In the future, we aim to improve \newg by integrating general system optimization techniques and expand its application areas to support more algorithm categories like machine learning algorithms and more graph types like dynamic graphs and temporal analytics.

%\end{document}  % This is where a 'short' article might terminate

% ensure same length columns on last page (might need two sub-sequent latex runs)

\balance

%ACKNOWLEDGMENTS are optional
\section{Acknowledgments}
\label{sec:ack}
Lu Qin is supported by ARC DE140100999 and DP160101513. Lijun Chang is supported by ARC DP160101513 and FT180100256. Ying Zhang is supported by ARC DE140100679 and DP170103710. Xuemin Lin is supported by NSFC61232006, ARC DP150102728, DP140103578 and DP170101628.

\bibliographystyle{abbrv}
\bibliography{sigproc}

\end{document}